\documentclass[twocolumn]{aastex63}
\usepackage[utf8]{inputenc}

\newcommand{\Msun}{$M_{\odot}$}
\newcommand{\HI}{\hbox{{\rm H}\kern 0.2em{\sc i}}}
\newcommand{\Hii}{\hbox{{\rm H}\kern 0.2em{\sc i}{\sc i}}}

\usepackage{amsmath}

\received{xxxx, 2025}
\revised{xxxx, 2025}
\accepted{xxxx, 2025}
\shorttitle{{\HI} turbulence in the SMC}
\shortauthors{Lee et al.}
\graphicspath{{./}{figures/}}

\begin{document}

\title{Study of {\HI} Turbulence in the SMC Using Multi-point Structure Functions}

\correspondingauthor{Bumhyun Lee, Min-Young Lee}
\email{bhlee301@gmail.com, mlee@kasi.re.kr}

\author[0000-0002-3810-1806]{Bumhyun Lee}
\affiliation{Department of Astronomy, Yonsei University, 50 Yonsei-ro, Seodaemun-gu, Seoul 03722, Republic of Korea}
\affiliation{Korea Astronomy and Space Science Institute, 776 Daedeokdae-ro, Daejeon 34055, Republic of Korea}

\author[0000-0002-9888-0784]{Min-Young Lee}
\affiliation{Korea Astronomy and Space Science Institute, 776 Daedeokdae-ro, Daejeon 34055, Republic of Korea}

\author[0000-0003-1725-4376]{Jungyeon Cho}
\affiliation{Department of Astronomy and Space Science, Chungnam National University, Daejeon, Republic of Korea}

\author[0000-0001-9504-7386]{Nickolas M. Pingel}
\affiliation{Department of Astronomy, University of Wisconsin-Madison, 475 North Charter St., Madison, WI, 53706-15821, USA}

\author[0000-0003-0742-2006]{Yik Ki Ma}
\affiliation{Max-Planck-Institut f\"ur Radioastronomie, Auf dem H\"ugel 69, 53121 Bonn, Germany}
\affiliation{Research School of Astronomy and Astrophysics, The Australian National University, Canberra, ACT 2611, Australia}

\author[0000-0001-7105-0994]{Katie Jameson}
\affiliation{Caltech Owens Valley Radio Observatory, Pasadena, CA 91125, USA}

\author[0000-0002-4899-4169]{James Dempsey}
\affiliation{Research School of Astronomy and Astrophysics, The Australian National University, Canberra, ACT 2611, Australia}

\author[0000-0002-9214-8613]{Helga D\'enes}
\affiliation{School of Physical Sciences and Nanotechnology, Yachay Tech University, Hacienda San Jos\'e S/N, 100119, Urcuqu\'{\i}, Ecuador}

\author[0000-0002-6300-7459]{John M. Dickey}
\affiliation{School of Natural Science, University of Tasmania, Hobart, TAS, Australia}

\author[0000-0002-0706-2306]{Christoph Federrath}
\affiliation{Research School of Astronomy and Astrophysics, The Australian National University, Canberra, ACT 2611, Australia}

\author[0000-0002-1495-760X]{Steven Gibson}
\affiliation{Department of Physics and Astronomy, Western Kentucky University, Bowling Green, KY 42101, USA}


\author[0000-0001-7462-4818]{Gilles Joncas}
\affiliation{D\'epartement de physique, de g\'enie physique et d'optique, Universit\'e Laval}

\author[0000-0002-6637-9987]{Ian Kemp}
\affiliation{International Centre for Radio Astronomy Research (ICRAR), Curtin University, Bentley, WA 6102, Australia}
\affiliation{CSIRO Space and Astronomy, 26 Dick Perry Avenue, Kensington, WA 6151, Australia}

\author[0000-0002-6760-7531]{Shin-Jeong Kim}
\affiliation{Research School of Astronomy and Astrophysics, The Australian National University, Canberra, ACT 2611, Australia}

\author[0000-0001-6846-5347]{Callum Lynn}
\affiliation{Research School of Astronomy and Astrophysics, The Australian National University, Canberra, ACT 2611, Australia}

\author[0000-0002-5501-232X]{Antoine Marchal}
\affiliation{Research School of Astronomy and Astrophysics, The Australian National University, Canberra, ACT 2611, Australia}

\author[0000-0003-2730-957X]{N. M. McClure-Griffiths}
\affiliation{Research School of Astronomy and Astrophysics, The Australian National University, Canberra, ACT 2611, Australia}

\author[0000-0002-2712-4156]{Hiep Nguyen}
\affiliation{Research School of Astronomy and Astrophysics, The Australian National University, Canberra, ACT 2611, Australia}

\author[0000-0001-9708-0286]{Amit Seta}
\affiliation{Research School of Astronomy and Astrophysics, The Australian National University, Canberra, ACT 2611, Australia}

\author[0000-0002-0294-4465]{Juan D. Soler}
\affiliation{University of Vienna, Department of Astrophysics, Türkenschanzstrasse 17, 1180 Vienna, Austria}
\affiliation{Istituto di Astrofisica e Planetologia Spaziali (IAPS). INAF. Via Fosso del Cavaliere 100, 00133 Roma, Italy}

\author[0000-0002-3418-7817]{Sne{\v z}ana Stanimirovi{\'c}}
\affiliation{University of Wisconsin–Madison, Department of Astronomy, 475 N Charter St, Madison, WI 53703, USA}

\author[0000-0002-1272-3017]{Jacco Th. van Loon}
\affiliation{Lennard-Jones Laboratories, Keele University, ST5 5BG, UK}





\begin{abstract}
Turbulence in the interstellar medium (ISM) plays an important role in many physical processes, including forming stars and shaping complex ISM structures. In this work, we investigate the {\HI} turbulent properties of the Small Magellanic Cloud (SMC) to reveal what physical mechanisms drive the turbulence and at what scales. Using the high-resolution {\HI} data of the Galactic ASKAP (GASKAP) survey and multi-point structure functions (SF), we perform a statistical analysis of {\HI} turbulence in 34 subregions of the SMC. Two-point SFs tend to show a linear trend, and their slope values are relatively uniform across the SMC, suggesting that large-scale structures exist and are dominant in the two-point SFs. On the other hand, seven-point SF enables us to probe small-scale turbulence by removing large-scale fluctuations, which is difficult to achieve with the two-point SFs. In the seven-point SFs, we find break features at scales of 34-84 pc, with a median scale of $\sim$50 pc. This result indicates the presence of small-scale turbulent fluctuations in the SMC and quantifies its scale. In addition, we find strong correlations between slope values of the seven-point SFs and the stellar feedback-related quantities (e.g., H$\alpha$ intensities, the number of young stellar objects, and the number of {\HI} shells), suggesting that stellar feedback may affect the small-scale turbulent properties of the {\HI} gas in the SMC. Lastly, estimated sonic Mach numbers across the SMC are subsonic, which is consistent with the fact that the {\HI} gas of the SMC primarily consists of the warm neutral medium. 
\end{abstract}





\section{Introduction} \label{sec:intro}
Turbulence in the interstellar medium (ISM) is thought to play a pivotal role in various physical processes such as the formation of molecular clouds, star formation, and shaping complex structures of the ISM (for reviews, see \citealt{elmegreen2004,mckee2007}). Depending on the driving scales, numerous mechanisms that drive ISM turbulence have been suggested. On small scales, stellar feedback, including stellar winds and supernova (SN) explosions, can be a power source for ISM turbulence \citep{kim2001,joung2006,grisdale2017}. In addition, thermal instabilities can be another energy source for maintaining ISM turbulence \citep{kritsuk2002,piontek2004}. Meanwhile, large-scale turbulence can be induced by galaxy interactions \citep{renaud2014}, gravitational instabilities \citep{wada2002,krumholz2016}, magneto-rotational instabilities \citep{sellwood1999,kim2003}, and gas accretions \citep{klessen2010}. Although previous studies have investigated various mechanisms for driving turbulence in the ISM, it is still unclear what physical process primarily characterizes the turbulent properties at a given spatial scale \citep{Elmegreen2009,FederrathEtAl2017iaus}. There are various statistical diagnostic approaches (see \citealt{elmegreen2004,boyden2016,koch2017}), such as the probability distribution functions (PDF) \citep{BruntFederrathPrice2010a,KainulainenFederrathHenning2014,bialy2017,burkhart2017}, spatial power spectrum (SPS) \citep{stanimirovic1999,FederrathKlessen2013,pingel2018}, structure function (SF) \citep{burkhart2015,szotkowski2019,SetaEtAl2023}, and delta variance \citep{stutzki1998,ossenkopf2008}, for probing the properties of turbulence \citep[see also][for a characterization of all of these techniques in numerical simulations]{FederrathDuvalKlessenSchmidtMacLow2010}.

The Small Magellanic Cloud (SMC) (stellar mass: 3.1~$\times$~10$^{8}$~{\Msun}; \citealt{skibba2012}), a dwarf irregular galaxy, is one of the nearest extragalactic neighbors, which is located at a distance of $\sim$60 kpc \citep{graczyk2020}. The SMC harbors a variety of turbulence-driving sources and modes \citep{gerrard2023}. As seen in Figure~\ref{fig:himap} and \ref{fig:ha_yso_shell}, many young stellar objects (YSOs) and active star-forming regions are present across the SMC. Their stellar feedback, such as outflows/jets of YSOs, stellar winds from massive stars, and SN explosions, can influence the ISM turbulent properties of the SMC at relatively small scales, from a few pc to $\sim$100 pc. Thanks to the close distance to the SMC, we can resolve and probe these YSOs and individual star-forming regions in detail. Compared to the Milky Way (MW), the SMC has different internal environments with low metallicity ($\sim$0.2~$Z_{\sun}$; \citealt{russell1992}) and a strong radiation field \citep{vangioni-flam1980, jameson2018, saldano2023}. The SMC has two interesting substructures (Figure~\ref{fig:himap}): i) the bar structure with relatively strong {\HI} emission and active star-forming sites; ii) the wing structure that is extended towards the Large Magellanic Cloud (LMC), which is thought to be produced by interactions with the LMC and the MW \citep{gordon2011,besla2012,kallivayalil2013,donghia2016}. In particular, the interaction with the LMC is thought to be one of the turbulence driving sources of the ISM at large scales \citep{goldman2000}. All of these characteristics make the SMC an interesting and unique laboratory to study ISM turbulence at a wide range of physical sizes, from the small scale ($\sim$pc scale) to the entire galactic scale ($\sim$kpc scale).

Several studies have examined the {\HI} turbulent properties in the SMC. In particular, analyzing the slope of the SPS or SF using {\HI} data allows us to study the spatial distribution of large- and small-scale {\HI} structures, their turbulent properties, and the mechanisms driving the turbulence. Two studies \citep{stanimirovic1999,stanimirovic2001} using the Australian Telescope Compact Array (ATCA) {\HI} data, combined with the Parkes telescope data, found that the {\HI} SPS of the SMC is characterized by a single power-law slope, suggesting that the driving scale of turbulence could be larger than the entire disk size of the SMC. With the same ATCA + Parkes {\HI} data (spatial resolution: $\sim$30~pc), \cite{nestingen-palm2017} probed whether the {\HI} turbulent properties are different between the central and outer regions of the SMC, using SF analysis. However, they did not find any significant differences and concluded that the {\HI} turbulent properties are uniform in the SMC at the 30 pc scale. Recently, \cite{szotkowski2019} investigated the spatial distribution of the SF and SPS slopes in the SMC using the {\HI} data (spatial resolution: $\sim$10~pc), obtained with commissioning observations \citep{mcclure2018} of the Australian Square Kilometre Array Pathfinder \citep[ASKAP;][]{hotan2021}. This study showed that there is no significant variation in slope values across the SMC, indicating similar turbulent properties. However, in the LMC, significant slope variations were found, with steep slopes near major {\Hii} regions due to stellar feedback \citep{szotkowski2019}.

Previous works have attempted to reveal what physical mechanisms are responsible for turbulence or at what scale it is driven in the SMC, but results are still inconclusive. One suggested mechanism is stellar feedback that could widely impact the entire SMC disk \citep[e.g.,][]{nestingen-palm2017}. Another possibility is that the interaction with the LMC and the MW \citep{donghia2016} induces the large-scale turbulence of the SMC \citep{goldman2000}, and then this turbulence could cascade down to the smaller scale \citep[e.g.,][]{stanimirovic2001,chepurnov2015}. 

In this work, we investigate the turbulent properties of the SMC in order to unravel the main drivers of turbulent energy injection and the scales at which they operate. Using new {\HI} data, obtained with the Galactic ASKAP (GASKAP) survey \citep{pingel2022}, we carried out a statistical analysis of {\HI} turbulence with multi-point structure functions \citep{cho2019,SetaFederrath2024}, primarily probing {\HI} density fluctuation by using a velocity-integrated {\HI} intensity map (see Section~\ref{sec:met}). The integrated {\HI} intensity is proportional to the column density under optically thin condition. We therefore use the integrated intensity as a measure of column density and consequently density fluctuations. In particular, the multi-point SFs can separate small- and large-scale fluctuations of the {\HI} density structures in the SMC by removing the large-scale fluctuations. This enables us to characterize the behavior of the small-scale turbulence in the SMC for the first time and constrain the origin of turbulence by utilizing the observational data related to the possible turbulence-driving sources (e.g., stellar feedback). 

In addition, the high-angular resolution (30$\arcsec$) of the GASKAP-{\HI} data allows us to probe small-scale structures of the SMC at $\sim$10 pc scale. Two recent studies using this new GASKAP-{\HI} data from the SMC reported new findings related to {\HI} turbulent properties. \cite{pingel2022} found an averaged shallower slope measured over the velocity range of 90$-$190 km~s$^{-1}$ compared to a previous study \citep{stanimirovic1999}. This suggests that the higher angular resolution of the new GASKAP-{\HI} data can resolve smaller scales, leading to more power at these scales. In addition, \cite{gerrard2023} found that the turbulence driving parameter (b) changes between 0.3 and 1.0 in the SMC; its median value is about 0.51, implying that the turbulence in the SMC is predominantly compressively driven. 

This paper is organized as follows. In Section~\ref{sec:data}, we describe the GASKAP-{\HI} data and the ancillary data related to stellar feedback. In Section~\ref{sec:met}, we introduce the multi-point structure functions as a tool for turbulence analysis in this work. In Section~\ref{sec:result}, we present results of two- and seven-point SF calculations for the entire disk and subregions of the SMC. In Section~\ref{sec:discu}, we discuss the {\HI} turbulent properties of the SMC. Section~\ref{sec:sum} summarizes our results and conclusions.

Throughout this paper, we adopted a distance of 60 kpc to the SMC \citep{graczyk2020}.

\section{Data} \label{sec:data}
\subsection{ASKAP + Parkes {\HI} Data} \label{sec:hi_data}

\begin{figure*}[!htbp]
\begin{center}
\includegraphics[width=1.0\textwidth]{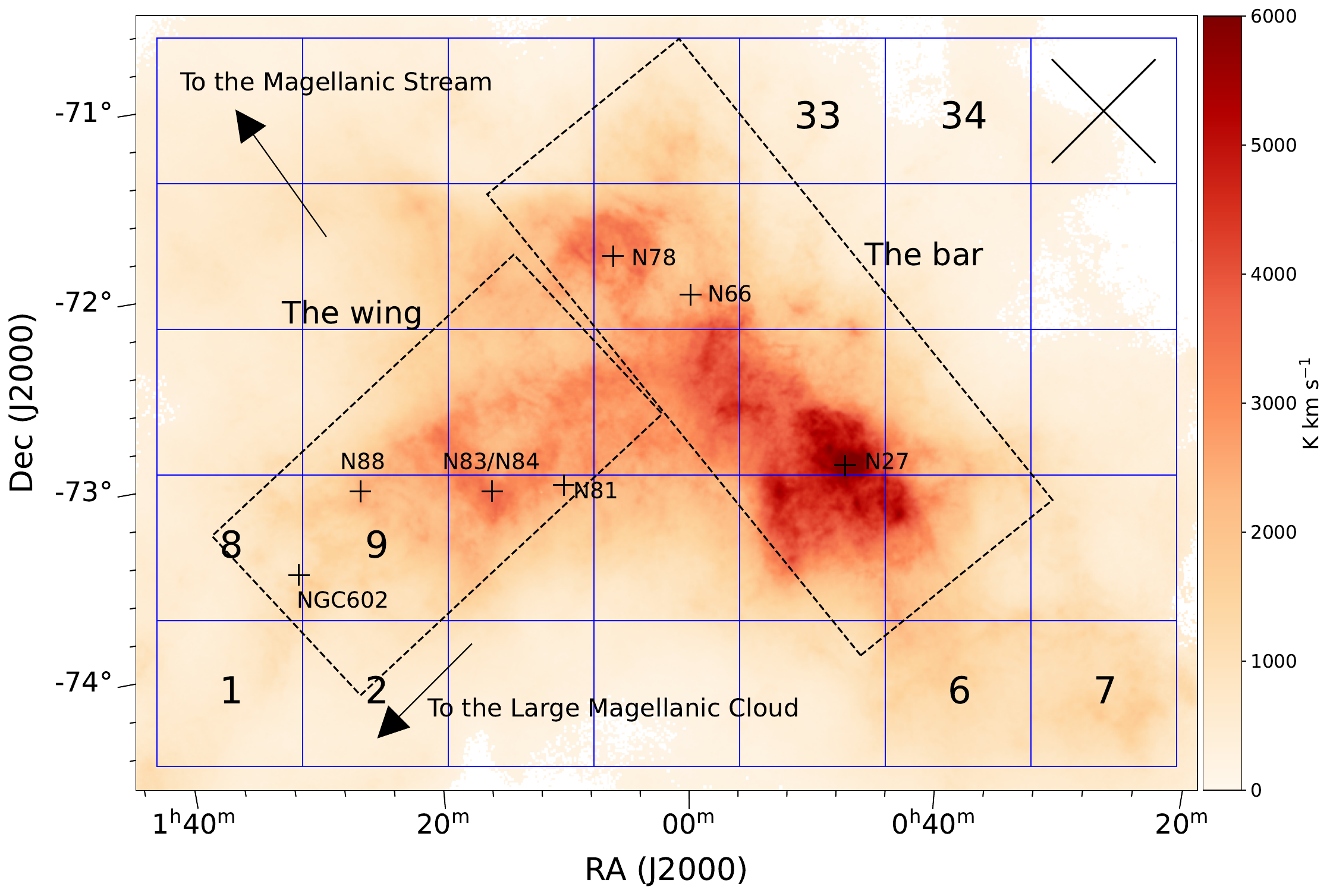}
\caption{GASKAP + Parkes {\HI} integrated intensity map of the SMC with a pixel size of 28$\arcsec$. Pixels (white) with S/N~$<$~5 are masked. The 35 subregions where we calculate structure functions are overlaid on the $W_{\rm HI}$ map. Each subregion (840~pc $\times$ 840~pc) is indicated by its corresponding number. Subregion 35 (marked with a large cross) was discarded in our analysis since more than 50\% of the pixels in this region have a low S/N ($<$5). The left and middle dashed rectangles show the Wing and the Bar locations, respectively. The directions to the Magellanic Stream and the Large Magellanic Cloud are indicated by two arrows \citep{mcclure2018,gerrard2023}. The small crosses are the representative star-forming regions in the SMC \citep[e.g.,][]{takekoshi2017, saldano2023}.} \label{fig:himap}
\end{center}
\end{figure*}

In this work, we used recent GASKAP-{\HI} pilot survey data \citep{pingel2022} combined with the Parkes single dish data from the Parkes Galactic All-Sky Survey (GASS; \citealt{mcclure2009, kalberla2010, kalberla2015}). The GASKAP-{\HI} pilot observations for the SMC were carried out with 36 antennas of the Australian Square Kilometre Array Pathfinder (ASKAP) in December 2019. The GASKAP-{\HI} data were calibrated using the ASKAPsoft \citep{guzman2019}. The final {\HI} data cube was created using a custom imaging pipeline with tasks from various radio data reduction packages: 1) WSClean for combining multiple ASKAP pointings onto a unified grid for joint deconvolution, and 2) the {\tt IMMERGE} task of {\tt MIRIAD} (\citealt{sault1995}) to combine the GASKAP-{\HI} data with the Parkes single dish data, addressing the missing flux issue due to a lack of short baselines. We refer the reader to \cite{pingel2022} for further details of the data reduction (calibration and imaging). The {\HI} cube that we used in this work covers $\sim$5$\degr$.7 $\times$ $\sim$4$\degr$.1 and contains 220 channels (43.8 km~s$^{-1}$ to 257.8 km~s$^{-1}$) with a velocity resolution of 0.98 km~s$^{-1}$. The synthesized beam size is 30$\arcsec$, and the original pixel size is 7$\arcsec$. The rms noise level in brightness temperature ($\sigma_{T}$) is 1.1~K over a channel width of 0.98 km~s$^{-1}$. We generated a velocity-integrated {\HI} intensity map (hereafter, the $W_{\rm HI}$ map) using a specific velocity range (57.5 km~s$^{-1}$ to 235.4 km~s$^{-1}$) showing {\HI} emission. This $W_{\rm HI}$ map was rebinned to make pixels have a similar angular size (28$\arcsec$$\approx$8.4 pc) to the beam size (30$\arcsec$$\approx$9.0 pc) of the GASKAP observations, which mitigates the correlation between pixels. The final $W_{\rm HI}$ map is shown in Figure~\ref{fig:himap}. We created 35 local {\HI} maps by dividing the entire $W_{\rm HI}$ map into 35 individual subregions to probe local {\HI} turbulent properties. Details of the subregions are described in Section~\ref{subsec:subregion}.

The rms noise levels in individual pixels of the binned {\HI} data cube with the pixel size of 28$\arcsec$ were measured using emission-free channels, and then we created a noise map using those measured rms noise levels from individual pixels. By combining this noise map and the peak {\HI} intensity map with the pixel size of 28$\arcsec$, we also generated a signal-to-noise (S/N) map. By inspecting the S/N map, we abandoned pixels with low S/N ($<$ 5) in our SF calculations. In particular, Subregion~35, which is marked with a large cross in Figure~\ref{fig:himap}, was discarded since more than 50\% of all pixels in this subregion have less than five-sigma detections.

\subsection{Ancillary Data Related to Stellar Feedback} \label{subsec:mul_data}

To probe any connections between the {\HI} turbulent properties and physical quantities related to stellar feedback in the individual subregions of the SMC (Figure~\ref{fig:himap}), we used the continuum-subtracted H$\alpha$ image from the Magellanic Cloud Emission Line Survey (MCELS; \citealt{smith1999}) and the information of YSOs and {\HI} shells obtained from previous works \citep{staveley-Smith1997, hatzidimitriou2005, sewilo2013}. 


\subsubsection{H$\alpha$ Data}
H$\alpha$ emission is commonly associated with massive young stars and {\Hii} regions. Stellar radiation and winds from the massive stars can be energy sources for affecting ISM turbulence \citep{gritschneder2009,walch2012,gallegos-garcia2020}. To compare with the {\HI} turbulent properties derived from the $W_{\rm HI}$ map with a pixel size of 28$\arcsec$, the H$\alpha$ image was convolved and regridded to match the pixel size (28$\arcsec$) of the {\HI} image. Then, we used the sum of H$\alpha$ intensities (in $\rm erg~s^{-1}~cm^{-2}$) as a measure of stellar feedback, which was estimated in each subregion of the SMC using the spatially matched H$\alpha$ image. The dust attenuation correction was not applied to the H$\alpha$ data, which could affect our analysis given the overall uncertainty of 20\% (\citealt{jameson2016}, private communication).

\subsubsection{Other Ancillary Data}
Outflows and jets of YSOs can also be responsible for driving and maintaining ISM turbulence in pc scale \citep{nakamura2011,narayanan2012}. \cite{sewilo2013} identified high-reliability YSO candidates in the SMC, based on the analysis for the color-magnitude diagrams and the spectral energy distribution fitting using the infrared (IR) source catalogs, IR and optical images of the SMC. In this work, we consider these candidates to represent the population of YSOs in the SMC.

{\rm H}\kern 0.2em{\sc i} shells are believed to be produced by (i) the stellar winds from massive stars and (ii) SN explosions \citep{staveley-Smith1997, kim1999, hatzidimitriou2005, park2013} that can inject energy into the ISM. The {\HI} shells in the SMC were identified from the position-velocity diagram of the ATCA {\HI} data \citep{staveley-Smith1997, hatzidimitriou2005}. \cite{staveley-Smith1997} initially created the SMC {\HI} shell catalog with 501 {\HI} shells, and \cite{hatzidimitriou2005} later added eight more, bringing the total to 509 shells. Furthermore, \cite{hatzidimitriou2005} found that $\sim$90\% of these shells are associated with young stellar counterparts, such as OB associations, star clusters, super giants, Wolf-Rayet stars, supernova remnants (SNRs). In addition to these young stellar counterparts, other physical mechanisms, such as Gamma rays and turbulence in the ISM \citep{elmegreen1997,perna2000}, are also suggested to form {\HI} shell-like structures, which may particularly account for the {\HI} shells with no energy sources in their central regions \citep{hatzidimitriou2005}. 

\begin{figure*}[!htbp]
\begin{center}
\includegraphics[width=1.0\textwidth]{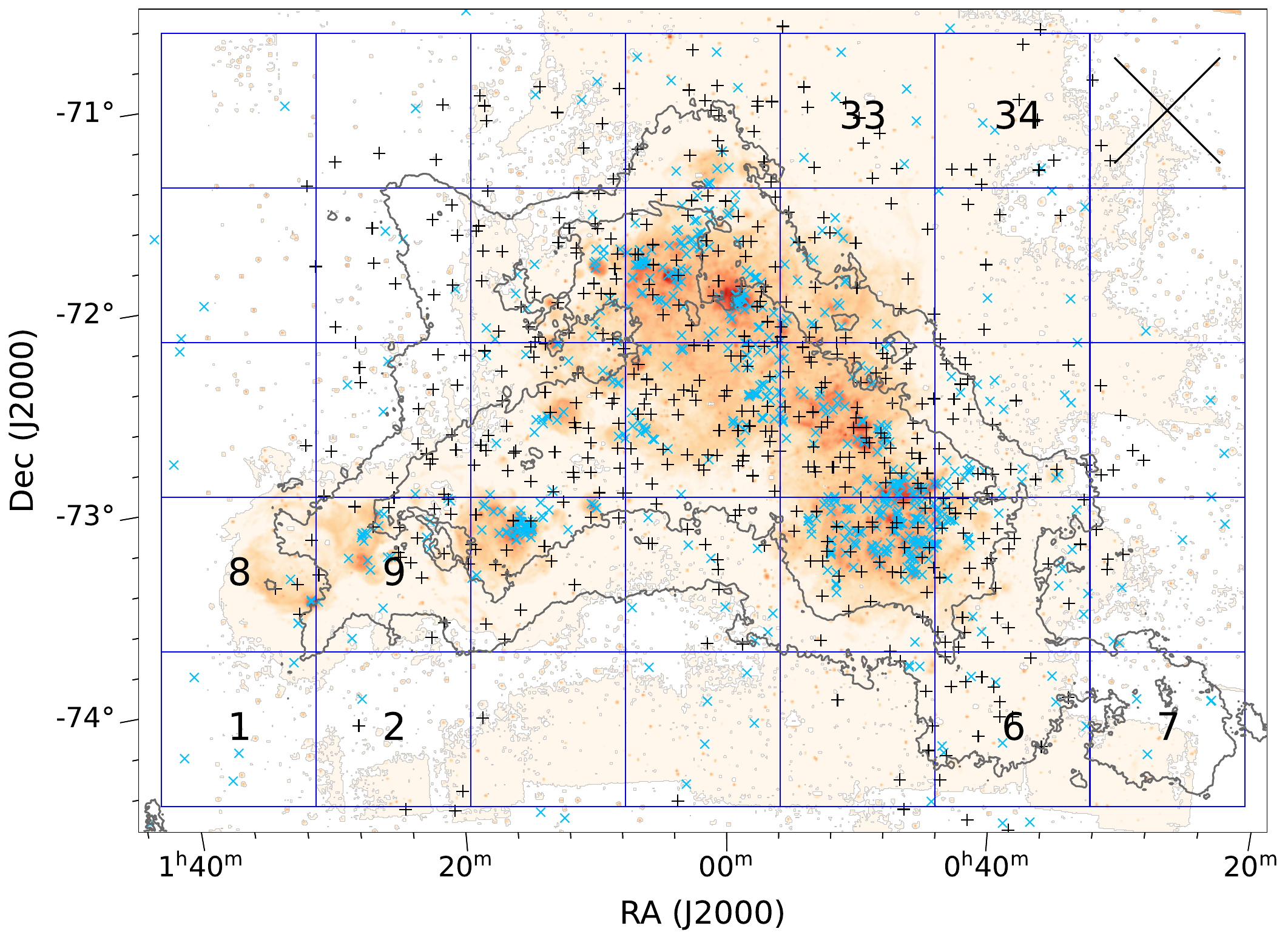}
\caption{YSOs (blue X mark) and {\HI} shell (black cross) distributions on the convolved H$\alpha$ image (color scale). The {\HI} column density distribution is shown in its {\HI} contour (gray) levels of 2 and 4 $\times$ 10$^{21}$ cm$^{-2}$. 
} \label{fig:ha_yso_shell}
\end{center}
\end{figure*}

In each subregion of the SMC, we counted the number of YSOs \citep{sewilo2013} and the number of {\HI} shells \citep{staveley-Smith1997,hatzidimitriou2005}. Figure~\ref{fig:ha_yso_shell} shows YSOs and {\HI} shell distributions on the H$\alpha$ image. Comparisons of the {\HI} turbulent properties with physical quantities related to stellar feedback (e.g., the sum of H$\alpha$ intensity value, the number of YSOs, and the number of {\HI} shells; these are what we use as a measure of stellar feedback) are presented in Section~\ref{subdiscus:stellar_feedback}.

\section{Methods} \label{sec:met}
To probe the turbulent properties of the {\HI} gas, we conducted a structure function (SF) analysis using the $W_{\rm HI}$ map. Since the SF calculations were performed on the image domain, it is relatively easy to interpret the results of the SF analysis, and we can avoid the Gibbs phenomenon (also known as the edge effect), which has been often observed in previous SPS studies (e.g., see Figure 6 of \citealt{pingel2013}). While the two-point SF has been conventionally applied for previous works, in this study, we utilized an improved approach with multi-point (e.g., three-, five-, seven-point) SFs, which helps us remove large-scale variations and obtain small-scale turbulent fluctuations \citep{cho2019}. Details of how multi-point SFs can remove large-scale fluctuations are described in \cite{cho2019} (see also Figure~1 of \citealt{cho2019}, and Figure~1 of \citealt{SetaFederrath2024}).

\subsection{Two-Point Structure Function} \label{sec:met2pt}
The two-point SF has been widely used in investigating the turbulent properties of the ISM \citep[e.g.,][]{KritsukEtAl2007,FederrathDuvalKlessenSchmidtMacLow2010,burkhart2015,nestingen-palm2017,szotkowski2019}. In our work, we calculated the two-point SF using the following equation \citep[][]{cho2019}: 

\begin{equation}
SF_{2pt}(r)~=~<\lvert I(\bold{x} + \bold{r}) - I(\bold{x})\rvert^2>,
\label{eq:st2pt}
\end{equation}

\noindent where $I(\bold{x})$ is the pixel value of the $W_{\rm HI}$ map at a given position $\bold{x}$, $\bold{r}$ is the distance from $\bold{x}$, and $r = \lvert \bold{r}\rvert$, and $\langle \rangle$ denotes average over the region.

\subsection{Multi-Point Structure Function} \label{sec:metnpt}
The variations in the $W_{\rm HI}$ map can arise from large- and small-scale density fluctuations. Regarding this, multi-point SF calculations\footnote{The n-point SF can remove a polynomial of degree n-2. For example, we can show that the three-point SF can remove a linear function $(ax + b)$ as follows: ${\Delta}I^{3pt} = I(x-r) - 2I(x) + I(x+r) = [a(x-r) + b] -2(ax+b) + [a(x+r) + b] = 0$. Using this and mathematical induction, we can prove that the n-point SF can remove a polynomial of degree n-2 (see the details in Appendix B of \citealt{cho2019}).} are greatly helpful to get rid of the influence of large-scale fluctuation. After removing the large-scale fluctuation, we can focus more on analyzing the turbulent properties related to the small-scale fluctuation. 

\begin{figure*}[!htbp]
\begin{center}
\includegraphics[width=1.00\textwidth]{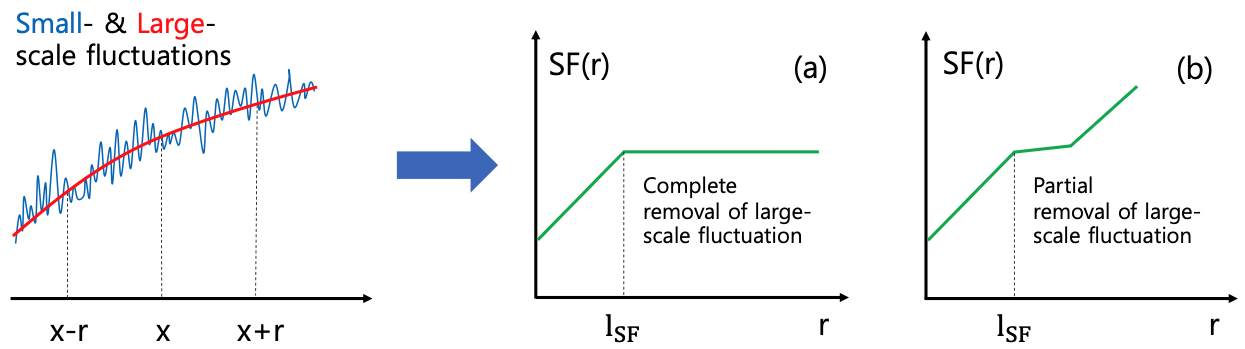} %
\caption{Cartoon showing various results of multi-point SF calculations for a complex structure, including small- and large-scale fluctuations. This image is reproduced based on Figure~1 of \cite{cho2019}. (a) In the case of complete removal of large-scale fluctuation (Case A), the multi-point SF becomes flat beyond the scale length of the small-scale fluctuation ($l_{\rm SF}$). (b) In the case of partial removal of large-scale fluctuation (Case B), the multi-point SF shows a break feature and keeps increasing afterwards.
\label{fig:sf_cartoon}}
\end{center}
\end{figure*}

If the large-scale variation is completely eliminated by subtracting the power of the large-scale variation using multi-point SFs (Case A$^{1}$), the SFs will remain flat beyond the characteristic spatial scale related to the small-scale fluctuation (Figure~\ref{fig:sf_cartoon} (a)). On the other hand, if the sampling size for the SF is larger than the typical scale of the large-scale variation, and/or if this variation has a complex shape, the large-scale variation is only partially removed (Case B). Then, the SFs will show break features briefly and keep increasing afterwards (Figure~\ref{fig:sf_cartoon} (b)). Same as the first case, the spatial scale at which the break feature of each SF emerges corresponds to the characteristic scale of the small-scale fluctuation. Hereafter, we call this spatial scale as the scale length of the small-scale fluctuation ($l_{\rm SF}$). 

In this work, we mainly used the seven-point SF, where a break feature is clearly visible, and it is defined as

\begin{equation}
    \begin{aligned}
        SF_{7pt}(r)~=&~\frac{1}{462}<\lvert I(\bold{x} - 3\bold{r}) - 6I(\bold{x} - 2\bold{r})\\
        &+ 15I(\bold{x} - \bold{r}) - 20I(\bold{x}) + 15I(\bold{x} + \bold{r})\\ 
        &- 6I(\bold{x} + 2\bold{r}) + I(\bold{x} + 3\bold{r})\rvert^2>,
    \end{aligned}
\label{eq:st7pt}
\end{equation}

\noindent \citep{cho2019, SetaFederrath2024}. We also calculated three-, four-, five-, six-point SFs and their results are presented in Appendix~\ref{app_multi_sf}. Our results show that the multi-point SF slopes appear to converge as the number of points increases (see Figure~\ref{fig:st234567_whole} in Appendix~\ref{app_multi_sf}), as recent studies using the multi-point SFs have reported \citep{SetaEtAl2023,SetaFederrath2024}.

When SF and SPS are fitted with power-law functions, i.e., $SF (r)$ $\propto$ $r^{-\alpha}$ and $SPS (k)$ $\propto$ $k^{-\beta}$, where $k$ is the wave number, there is an approximate linear relationship between the two indices (i.e.,  $\alpha = \beta - 2$, if $2 < \beta < 4$; \citealt{simonetti1984}). Therefore, we can directly compare the slope ($\alpha$) of SFs in our analysis with that of SPS ($\beta$) in other previous studies.

\subsection{Uncertainties in Structure Functions } \label{subsec:error}
For an error estimation of SFs, we employed a Monte Carlo approach by following \cite{pingel2013,pingel2018}. Specifically, we generated 1,000 simulated $W_{\rm HI}$ maps. For every simulated map, we set the value of each pixel by drawing a single sample from a Gaussian distribution whose mean equals the original intensity value at the given pixel and whose standard deviation matches its 1$\sigma$ uncertainty. For this Monte Carlo simulation, we adopted the standard deviation determined by $\sigma \times \sqrt{N_{chan}} \times \Delta v$, where $\sigma$ is the rms noise, $N_{chan}$ is the number of channels used for generating the $W_{\rm HI}$ map and $\Delta v$ is the velocity resolution. Then, we calculated SFs for the 1000 $W_{\rm HI}$ maps and took the standard deviation of the 1000 SFs as the uncertainty of each SF. The error range of the SFs is two to five orders of magnitude lower than the power of the SFs at a given spatial scale. Thus, most error bars of our SFs are extremely small.

\subsection{Subregions of the SMC} \label{subsec:subregion}
Various mechanisms driving turbulence can have different driving scales with a large range \citep{TamburroEtAl2009,BruntHeyerMacLow2009,Elmegreen2009,FederrathEtAl2017iaus}. For example, while the interaction with the LMC and the Milky Way can affect a large-scale region of the SMC, stellar feedback from active star-forming regions can drive turbulence on small scales. These different driving mechanisms are expected to make a difference in the turbulent properties in a given region and at a given spatial scale. In particular, the complex substructures (e.g., the wing) of the SMC (Figure~\ref{fig:himap}) and the inhomogeneous distribution of star-forming regions across the SMC (Figure~\ref{fig:ha_yso_shell}) may cause local variations in the turbulent properties. To investigate whether the local {\HI} turbulent properties change across subregions of the SMC, we divided the entire region of the SMC into 35 subregions (see Figure~\ref{fig:himap}). Each subregion with the size of 840 pc $\times$ 840 pc contains several star-forming sites with some {\HI} shells and YSOs, allowing us to study the influence of stellar feedback on {\HI} turbulence. It also has a sufficient number of pixels ($\sim$10000 pixels) to calculate the SFs properly. In particular, Subregion 35, which is marked with a large cross in several Figures (e.g., Figure~~\ref{fig:himap} and Figure~\ref{fig:sub2pt}), was discarded since more than 50\% of the pixels in this subregion have less than the five-sigma detection.

\section{Results} \label{sec:result}
\subsection{A Whole Region of the SMC} \label{subsec:wholesmc}

\begin{figure}[!htbp]
\begin{center}
\includegraphics[width=0.47\textwidth]{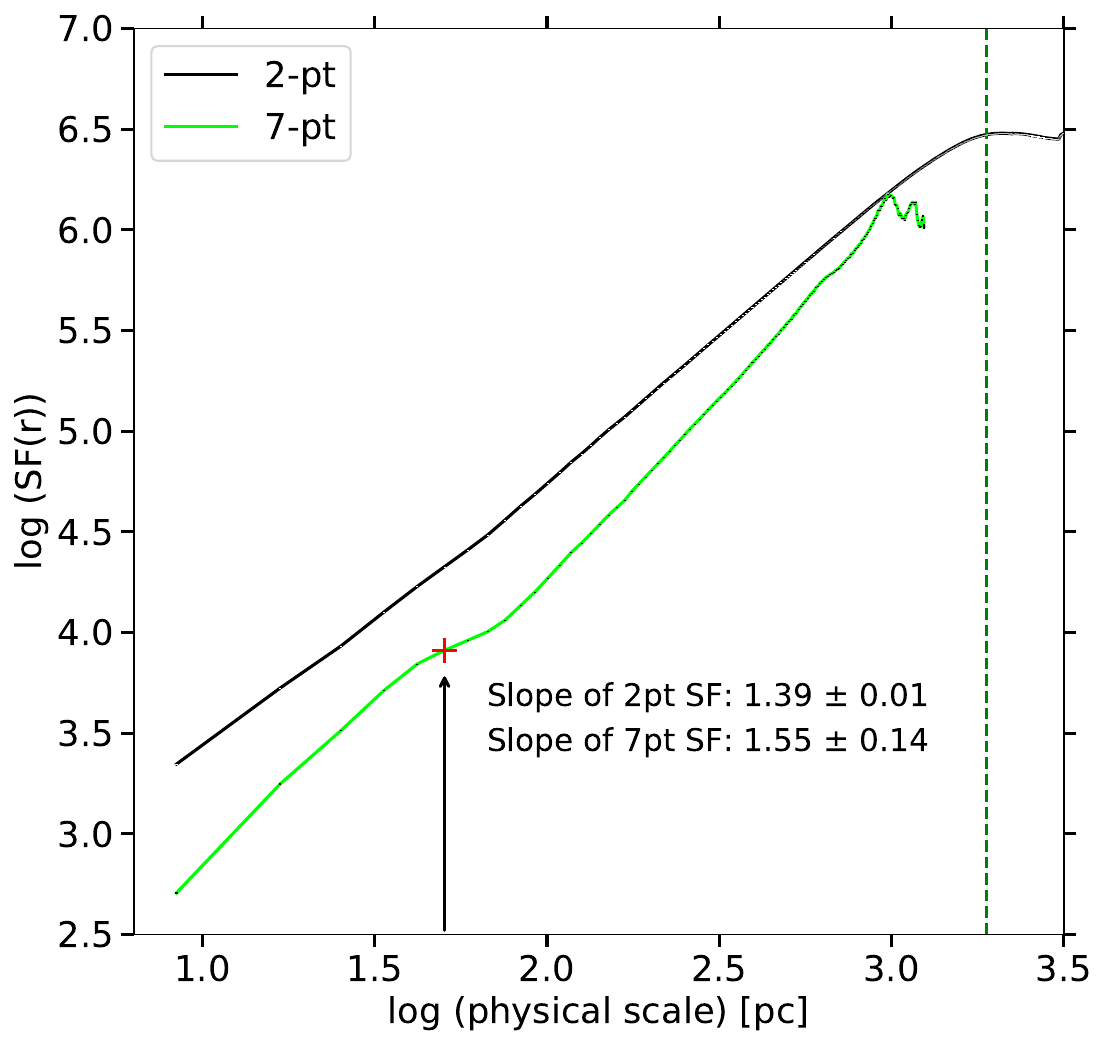} %
\caption{
Two- and seven-point SFs for the entire $W_{\rm HI}$ map of the SMC. Two- and seven-point SFs are shown in black and green, respectively. In the log-log plane, while the two-point SF is linear, the seven-point SF show a break (red cross) at a spatial scale of $\sim$50 pc. The slope of the two-point SF was estimated in the range of 8 pc to 1890 pc, just before the appearance of the turnover (green dashed line). Its slope value is 1.39 $\pm$ 0.01. The slope of the seven-point SF estimated over small scales (8$-$50 pc) is 1.55~$\pm$~0.14.
}

\label{fig:wholesf}
\end{center}
\end{figure}

Using Equations~(\ref{eq:st2pt}) and (\ref{eq:st7pt}), we calculated the two- and seven-point SFs for the entire $W_{\rm HI}$ map of the SMC; the results are shown in Figure~\ref{fig:wholesf}. Note that the sharp drops in the SFs at large spatial scales ($>$~1~kpc) are due to the limited map size \citep{hou1998, szotkowski2019}. These sharp drop features are also seen in the SFs of the subregions (Figure~\ref{fig:sub7pt}).

To estimate the slope value of each SF and its uncertainty, we used the jackknife method for resampling by following the approach of \cite{pingel2013}. In each resampling step, we excluded one data point from the original SF with $n$ data points and conducted non-weighted linear fitting to the subsample of $(n - 1)$ data points. We performed this process $n$ times. Then, with the derived slopes and their errors, the mean and the standard deviation were calculated as the final slope value and its uncertainty. 

As seen in Figure~\ref{fig:wholesf}, the two-point SF (black solid line) is roughly a straight line in the log-log plot. We estimated the slope of the two-point SF using the spatial scale ranges up to the point where the sharp drop appears. The slope value estimated over a spatial range of 8$\sim$1890~pc is 1.39~$\pm$~0.01, which is marginally consistent, within 3$\sigma$, with the slope value (1.25 $\pm$ 0.04) estimated by \cite{nestingen-palm2017}. In contrast to the linear trend of the two-point SF, we found a break feature at a spatial scale of $\sim$50 pc by visual inspection in the seven-point SF (green line in Figure~\ref{fig:wholesf}). This result indicates that the {\HI} disk of the SMC has small-scale fluctuations at $\sim$50 pc scale. We estimated the slope of the seven-point SF over the small-scale range up to the point where the break appears. Its slope value estimated over a smaller range of 8$\sim$50~pc is 1.55~$\pm$~0.14, which is higher than the slope of the two-point SF. 


\subsection{Subregions of the SMC} \label{subsec:subsmc}
\begin{figure*}[!htbp]
\begin{center}
\includegraphics[width=1.00\textwidth]{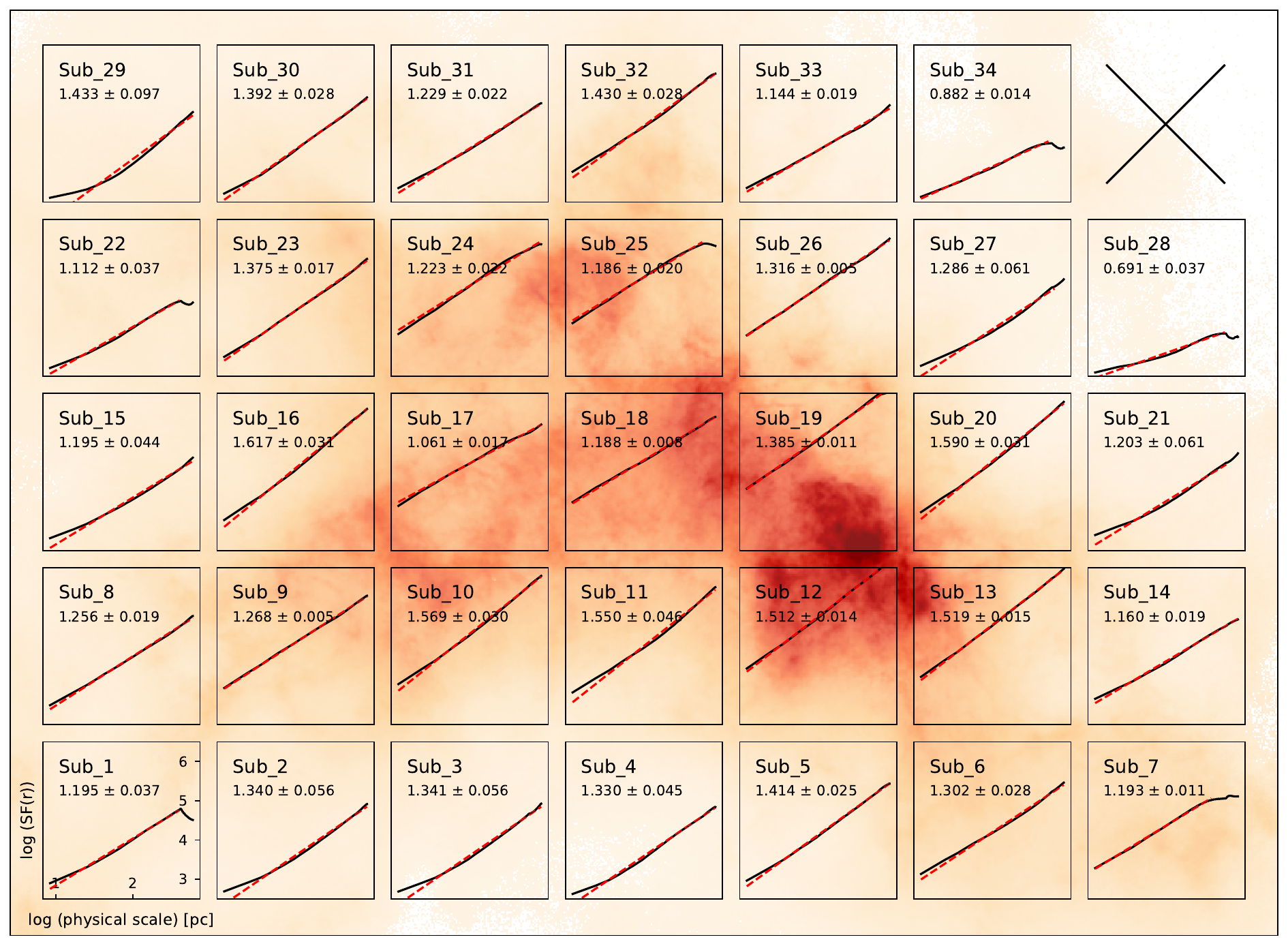} %
\caption{Results of the two-point SFs (black line) for the 34 subregions, overlaid on the $W_{\rm HI}$ map of the SMC. Their error bars are shown in black, but most of the error bars are quite small. The linear fit (red dashed lines) was performed in each subregion. The slope values (black) estimated from the two-point SFs range from $\sim$0.69 to $\sim$1.62.
\label{fig:sub2pt}}
\end{center}
\end{figure*}

\begin{figure*}[!htbp]
\begin{center}
\includegraphics[width=1.00\textwidth]{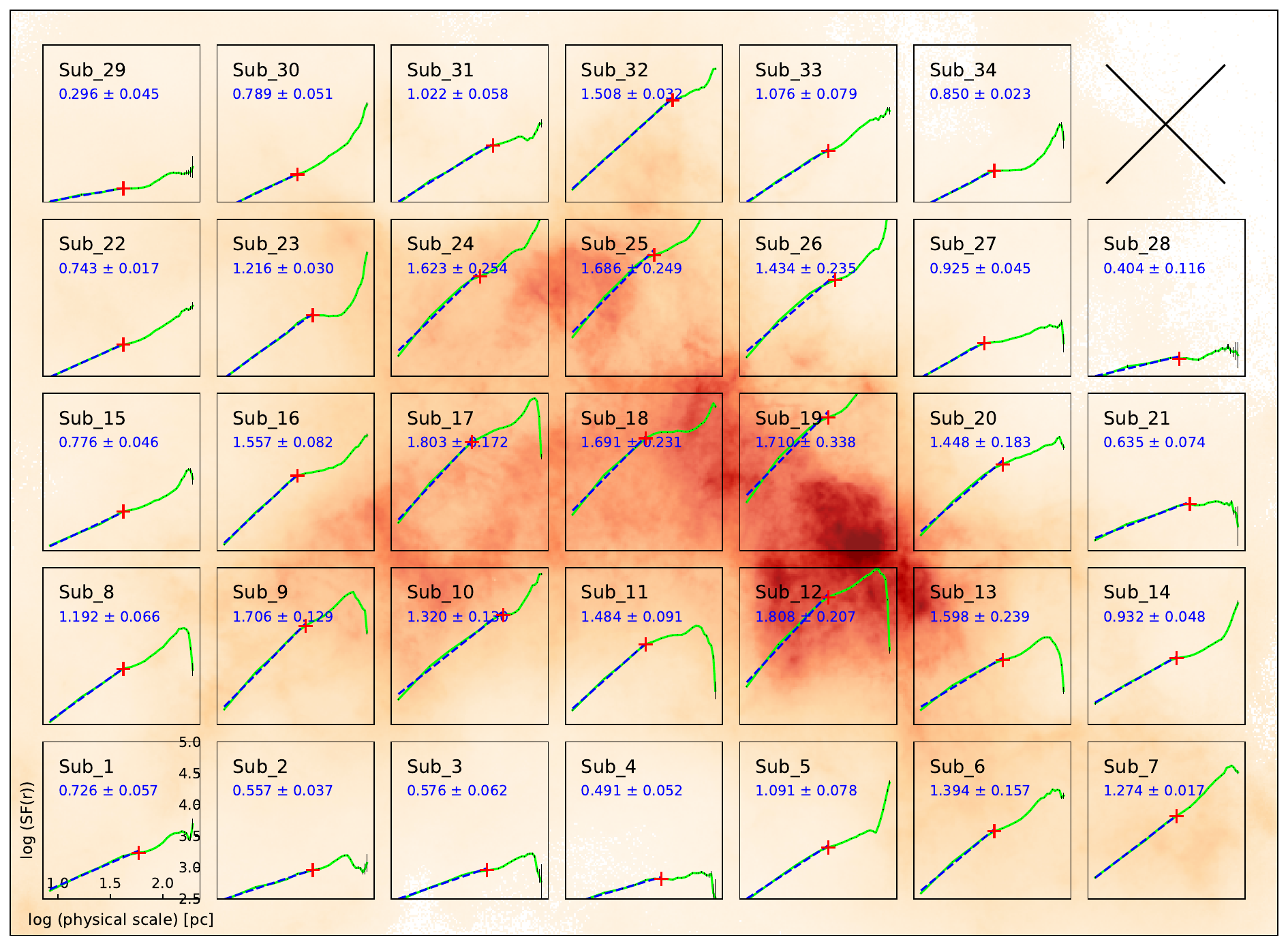} %
\caption{Results of the seven-point SFs (green line) for the 34 subregions, overlaid on the $W_{\rm HI}$ map of the SMC. Their error bars are shown in black, but most of the error bars are quite small. The seven-point SFs show the break feature (marked with a red cross) at between $\sim$34 pc and $\sim$84 pc. The linear fit (blue dashed line) was performed for each subregion over the spatial scale ranging from 8.4 pc up to the point where the break appears. The slope values (blue) estimated from the seven-point SFs range from 0.30 to 1.81.
\label{fig:sub7pt}}
\end{center}
\end{figure*}

In addition to probing the SFs for the entire SMC, we calculated the multi-point SFs for the 34 subregions (see Figure~\ref{fig:himap}) of the SMC to investigate whether the turbulent properties change across the SMC. Figures~\ref{fig:sub2pt} and \ref{fig:sub7pt} show two- and seven-point SFs (shown as black and green lines, respectively) in the subregions. In calculating the slopes of the two- and seven-point SFs for the SMC subregions, we also used jackknife resampling, as in the previous section. 

The two-point SFs are also linear in most subregions, similar to the two-point SF of the entire {\HI} disk of the SMC. The slopes of the two-point SFs in the subregions were estimated over spatial scales ranging from 8 pc to 353$-$596 pc. The slope values estimated from the two-point SFs of the subregions range from $\sim$0.69 to $\sim$1.62 (Figure~\ref{fig:sub2pt} and Table~\ref{tab:tur_vals}), and their median is 1.29, which is lower than the slope value (1.39) estimated using the entire {\HI} map.

In contrast to the behavior of the two-point SFs, the seven-point SFs show the break features in the subregions as seen in Figure~\ref{fig:sub7pt}. These features are found between $\sim$34 pc and $\sim$84 pc, and their median is 50 pc. The slopes of the seven-point SFs were calculated over spatial scales ranging from 8 pc to 34$-$84~pc. The slope values are between $\sim$0.30 and $\sim$1.81 (Figure~\ref{fig:sub7pt} and Table~\ref{tab:tur_vals}). The slope values of the seven-point SFs tend to be higher in {\HI}-bright regions (e.g., Subregion 12 and 19) compared to the outer regions with faint {\HI} emission (e.g., Subregion 2 and 21). The median of the slope values for the subregions is 1.20, which is somewhat different from the slope (1.55) estimated from the entire {\HI} map. The slope value from the entire SMC is close to the slope values of the subregions with relatively strong {\HI} emission (e.g., subregions 11 and 12). It may imply that the subregions with high pixel values contribute more to the seven-point SF calculations for the entire {\HI} map of the SMC. We will discuss variations in the slope values of the two- and seven-point SF in the SMC in Section~\ref{subsec:diff_behaviors}.

\begin{deluxetable*}{lccccc}
\tablecaption{Derived physical values from the entire region and individual subregions of the SMC using the multi-point SFs \label{tab:tur_vals}}
\tablehead{
\colhead{Region} & \colhead{Slope of 2pt SF} & \colhead{Slope of 7pt SF} & \colhead{$l_{\rm SF}$ (pc)} & \colhead{Mach number} & \colhead{Num. pixels}
\\
\colhead{(1)} & \colhead{(2)} & \colhead{(3)} & \colhead{(4)} & \colhead{(5)} & \colhead{(6)}
}
\startdata
Entire & 1.39$\pm$0.01 & 1.55$\pm$0.14 & 50  & -  & - \\
Sub\_1 & 1.195$\pm$0.037 & 0.726$\pm$0.057 & 59 & 0.45 & 10000 \\
Sub\_2 & 1.340$\pm$0.056 & 0.557$\pm$0.037 & 59 & 0.35 & 10000 \\
Sub\_3 & 1.341$\pm$0.056 & 0.576$\pm$0.062 & 59 & 0.50 & 9665 \\
Sub\_4 & 1.330$\pm$0.045 & 0.491$\pm$0.052 & 59 & 0.54 & 9667 \\
Sub\_5 & 1.414$\pm$0.025 & 1.091$\pm$0.078 & 50 & 0.43 & 10000 \\
Sub\_6 & 1.302$\pm$0.028 & 1.394$\pm$0.157 & 42 & 0.32 & 10000 \\
Sub\_7 & 1.193$\pm$0.011 & 1.274$\pm$0.017 & 50 & 0.47 & 10000 \\
Sub\_8 & 1.256$\pm$0.019 & 1.192$\pm$0.066 & 42 & 0.46 & 10000 \\
Sub\_9 & 1.268$\pm$0.005 & 1.706$\pm$0.129 & 50 & 0.45 & 10000 \\
Sub\_10 & 1.569$\pm$0.030 & 1.320$\pm$0.130 & 84 & 0.47 & 10000 \\
Sub\_11 & 1.550$\pm$0.046 & 1.484$\pm$0.091 & 59 & 0.34 & 10000 \\
Sub\_12 & 1.512$\pm$0.014 & 1.808$\pm$0.207 & 50 & 0.38 & 10000 \\
Sub\_13 & 1.519$\pm$0.015 & 1.598$\pm$0.239 & 50 & 0.40 & 10000 \\
Sub\_14 & 1.160$\pm$0.019 & 0.932$\pm$0.048 & 50 & 0.57 & 9992 \\
Sub\_15 & 1.195$\pm$0.044 & 0.776$\pm$0.046 & 42 & 0.45 & 9967 \\
Sub\_16 & 1.617$\pm$0.031 & 1.557$\pm$0.082 & 42 & 0.33 & 10000 \\
Sub\_17 & 1.061$\pm$0.017 & 1.803$\pm$0.172 & 59 & 0.34 & 10000 \\
Sub\_18 & 1.188$\pm$0.008 & 1.691$\pm$0.231 & 59 & 0.29 & 10000 \\
Sub\_19 & 1.385$\pm$0.011 & 1.710$\pm$0.338 & 50 & 0.38 & 10000 \\
Sub\_20 & 1.590$\pm$0.031 & 1.448$\pm$0.183 & 50 & 0.48 & 9980 \\
Sub\_21 & 1.203$\pm$0.061 & 0.635$\pm$0.074 & 67 & 0.67 & 9843 \\
Sub\_22 & 1.112$\pm$0.037 & 0.743$\pm$0.017 & 42 & 0.42 & 10000 \\
Sub\_23 & 1.375$\pm$0.017 & 1.216$\pm$0.030 & 59 & 0.34 & 10000 \\
Sub\_24 & 1.223$\pm$0.022 & 1.623$\pm$0.254 & 50 & 0.38 & 10000 \\
Sub\_25 & 1.186$\pm$0.020 & 1.686$\pm$0.249 & 50 & 0.45 & 10000 \\
Sub\_26 & 1.316$\pm$0.005 & 1.434$\pm$0.235 & 59 & 0.64 & 10000 \\
Sub\_27 & 1.286$\pm$0.061 & 0.925$\pm$0.045 & 34 & 0.64 & 9939 \\
Sub\_28 & 0.691$\pm$0.037 & 0.404$\pm$0.116 & 50 & 0.96 & 6117 \\
Sub\_29 & 1.433$\pm$0.097 & 0.296$\pm$0.045 & 42 & 0.45 & 9995 \\
Sub\_30 & 1.392$\pm$0.028 & 0.789$\pm$0.051 & 42 & 0.35 & 10000 \\
Sub\_31 & 1.229$\pm$0.022 & 1.022$\pm$0.058 & 67 & 0.69 & 9980 \\
Sub\_32 & 1.430$\pm$0.028 & 1.508$\pm$0.032 & 76 & 0.79 & 10000 \\
Sub\_33 & 1.144$\pm$0.019 & 1.076$\pm$0.079 & 50 & 0.83 & 9809 \\
Sub\_34 & 0.882$\pm$0.014 & 0.850$\pm$0.023 & 42 & 1.02 & 8899 \\
\enddata
\tablecomments{(1) region of the SMC; (2) slope value of the two-point SF; (3) slope value of the seven-point SF; (4) the scale length of the small-scale fluctuation ($l_{\rm SF}$); (5) Sonic Mach number for the small scales, derived using the results of the seven-point SFs; (6) the number of pixels with S/N $>$5, which are used to calculate the structure functions}
\end{deluxetable*}

\subsection{Sonic Mach Number} \label{subsec:mach_num}
The variance of density fluctuations can be directly linked to the sonic Mach number ($M_{S}$) (i.e., the 3D density variance$-$$M_{S}$ relation) when turbulence is the dominant mechanism for shaping the structures of the ISM \citep{FederrathKlessenSchmidt2008,FederrathDuvalKlessenSchmidtMacLow2010,burkhart2010,burkhart2012}. Furthermore, this relationship is even valid for 2D column density variance (i.e., the 2D column density variance$-$$M_{S}$ relation), as expressed in the following equation \citep{burkhart2012}: 

\begin{equation}
\sigma^{2}_{\Sigma/\Sigma_{0}} = (b^{2}M^{2}_{s} + 1)^{A} - 1,
\label{eq:mach_1}
\end{equation}

\noindent where $\sigma_{\Sigma/\Sigma_{0}}$ is the standard deviation of the 2D column density divided by its mean value, $b$ is the turbulence-driving parameter \citep{FederrathKlessenSchmidt2008,FederrathDuvalKlessenSchmidtMacLow2010,gerrard2023}, and $A$ is the specific scaling parameter \citep{burkhart2012}. 

In our SF analysis, the values of the seven-point SFs at the spatial scales where the break or flat feature appears are equal to the standard deviation of the small-scale fluctuations \citep{cho2019}. Thus, we can calculate the sonic Mach number using the following equation, which is rewritten based on Equation~\ref{eq:mach_1}: 

\begin{equation}
M_{s} = \sqrt{\frac{(\sigma^{2}_{\Sigma/\Sigma_{0}} + 1)^{1/A} - 1}{b^{2}}}.
\label{eq:mach_2}
\end{equation}

\begin{figure*}[!htbp]
\begin{center}
\includegraphics[width=1.0\textwidth]{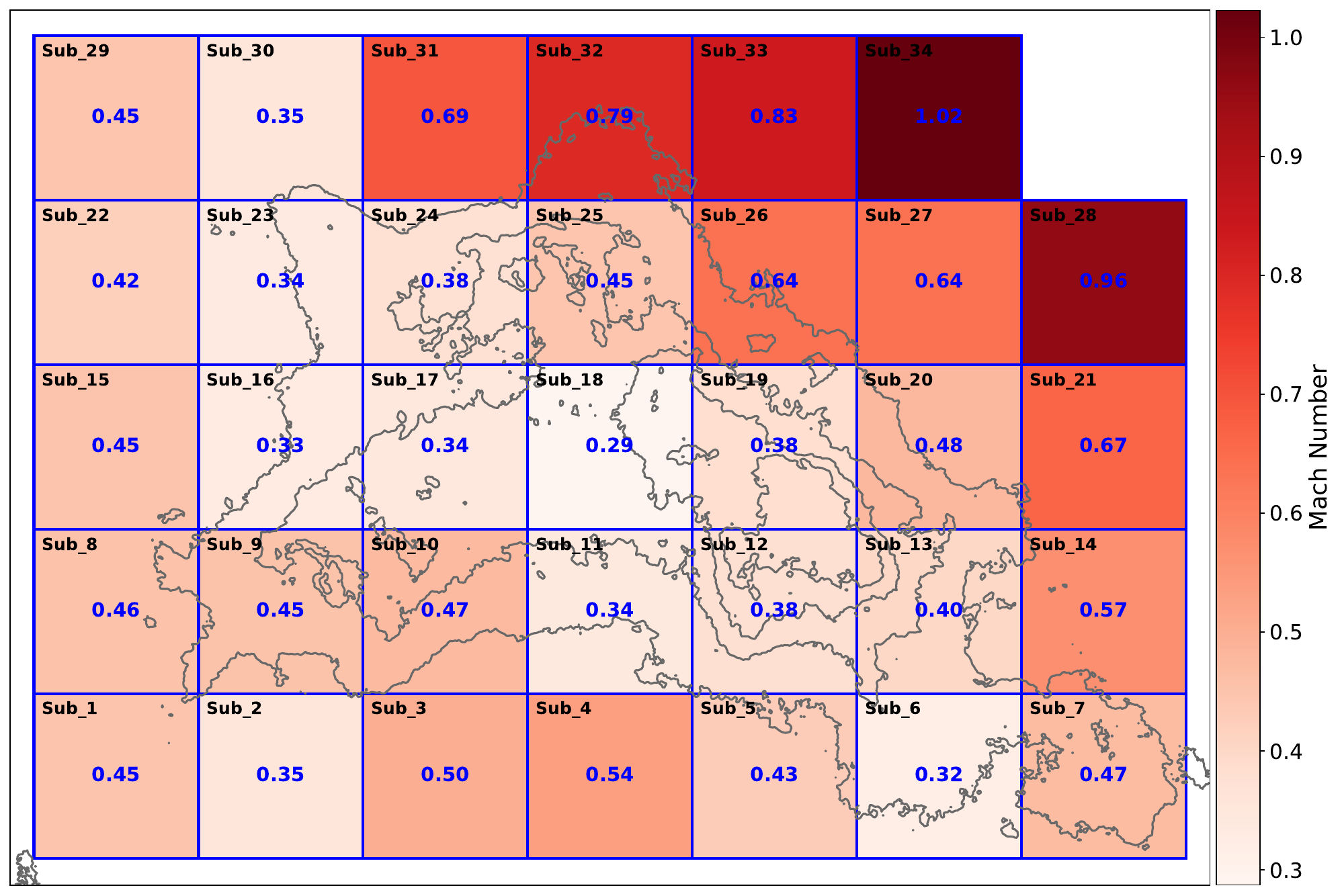} %
\caption{Sonic Mach numbers (red color scale and blue) in the individual subregions of the SMC. The {\HI} column density is shown in its {\HI} contour (gray) levels of 2, 4, 6, and 8 $\times$ 10$^{21}$ cm$^{-2}$. 
\label{fig:mach_num}}
\end{center}
\end{figure*}

\noindent Figure~\ref{fig:mach_num} shows sonic Mach numbers for small-scales in the individual subregions of the SMC, which were calculated by assuming specific values of the two parameters\footnote{Under the solenoidal driving condition ($b = 1/3$), \cite{burkhart2012} found the best-fit value of $A =0.11$ from the column density variance$-$$M_{s}$ relationship (Figure~2 of \citealt{burkhart2012}) using their MHD simulations. See also \cite{yoon2024}.} ($A = 0.11$ and $b = 1/3$; \citealt{burkhart2012}) and using the normalized standard deviation of the integrated intensity values ($\sigma^{2}_{\Sigma/\Sigma_{0}}$) in each subregion. Note that our calculation for Mach numbers is appropriate for an incompressible driving condition with a driving scale of $\sim$50 pc. Sonic Mach numbers across the SMC range from 0.32 to 1.02, indicating that all subregions are subsonic to transonic. We also found relatively high $M_{s}$ ($>$ 0.6) at the northwest region of the SMC. We will discuss sonic Mach numbers of the SMC in more detail in Section~\ref{subsec:mach_num}.

\section{Discussion} \label{sec:discu}

\subsection{{\HI} Turbulence Properties: Different Behaviors between Two- and Seven-point SFs} \label{subsec:diff_behaviors}
In this section, we discuss different behaviors between two- and seven-point SFs in the SMC and what physical mechanisms could be related to the variations of the slope values in the two- and the seven-point SFs. 

As described in Section~\ref{sec:result}, the two- and seven-point SFs show different behaviors in the SMC. The two-point SFs, which reflect contributions from both large- and small-scale structures, are found to be linear in many subregions of the SMC. With the exception of Subregion 28 and 34, showing relatively low slope values ($\sim$0.69 and $\sim$0.88), the slope values across the {\HI} disk of the SMC ($\sim$1.1$-$1.6) do not vary significantly. These values are also in generally good agreement with the range of SF slopes ($\sim$1.0$-$1.6) reported in the previous study \citep{szotkowski2019}. 

On the other hand, the slope values of the seven-point SFs, which represent exclusively the physical properties of the small-scale structures, change relatively more than the slope values of the two-point SFs. A relatively high standard deviation ($\sigma_{7pt}$: 0.45) of the slope values of the seven-point SFs, compared to the standard deviation ($\sigma_{2pt}$: 0.19) of the slope values of the two-point SFs, also supports our results that the slope values of the seven-point SFs vary widely compared to those of the two-point SFs. 

In particular, to explain the behavior (linear trend in many subregions and relatively uniform slope values across the SMC) of the two-point SF in the SMC, several hypotheses have been suggested. One scenario is that large-scale turbulence can govern the overall shape of the two-point SFs. Using magnetohydrodynamic (MHD) simulations, \cite{yoo2014} probed the turbulence driven by injecting energy at the same time on small- and large-spatial scales. They found that energy injection on the large-spatial scale can make large-scale fluctuations dominant and result in the change of the slope of the column density spectrum. Based on this result, the interactions between the SMC, LMC and the MW \citep{donghia2016}, which are larger than the SMC scale, can be one of the possible mechanisms that provide energy driving the large-scale turbulence in the SMC \citep{goldman2000, nestingen-palm2017,szotkowski2019} and may determine the overall shape of the two-point SFs in the SMC. 

In contrast to the behavior of the two-point SFs, the larger variation in the slope values of the seven-point SFs suggests that the small-scale fluctuations in the SMC seem to be relatively more affected by local processes. The small-scale structures could be shaped by turbulence driven by stellar feedback, which can affect the slope of the SF (either steeper or flatter)\citep{kowal2007,burkhart2010, walker2014, grisdale2017}. We will discuss the effect of stellar feedback on the slope of SFs in the next Section (Section~\ref{subdiscus:stellar_feedback}).

\begin{figure*}[!htbp]
\begin{center}
\includegraphics[width=1.00\textwidth]{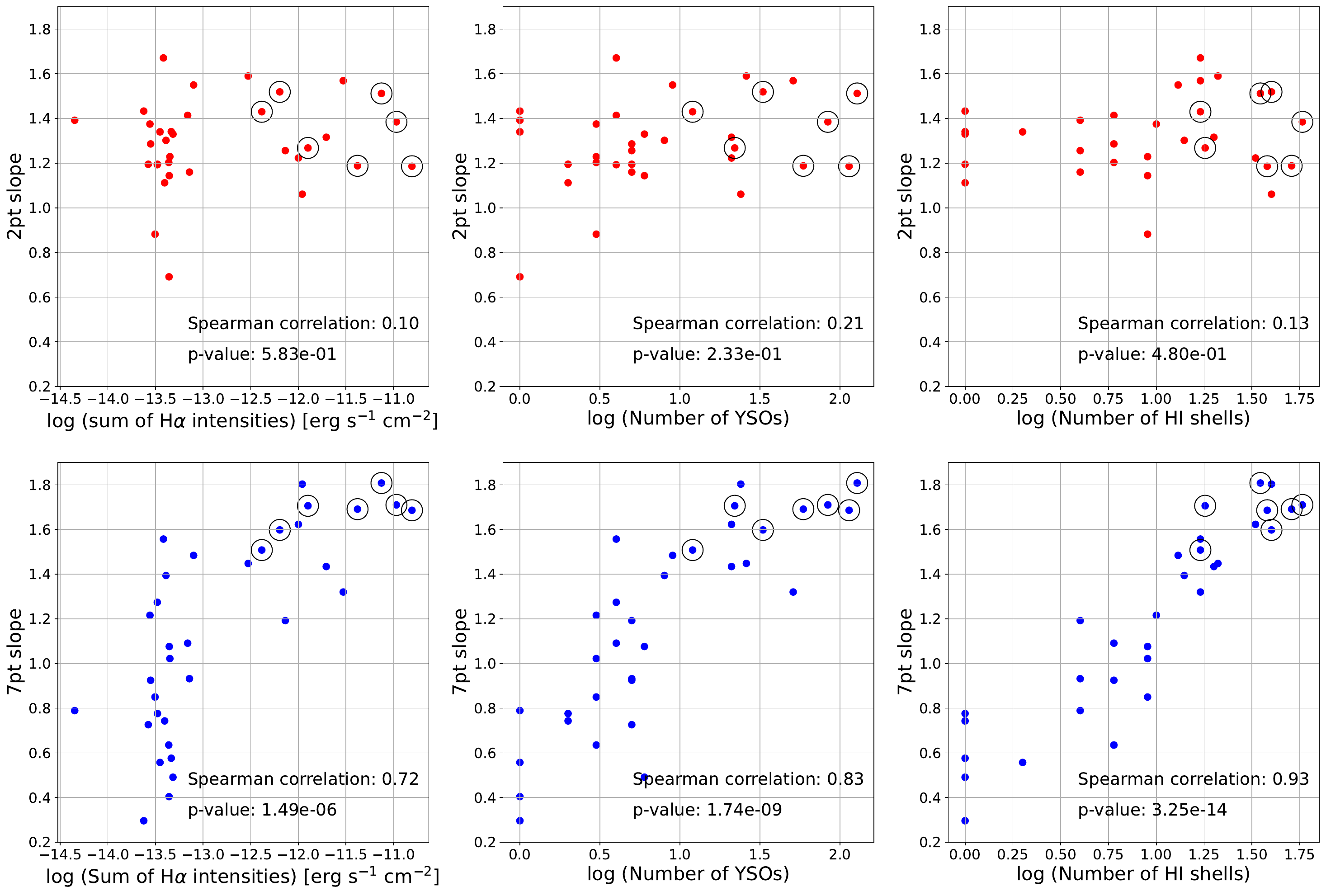} %
\caption{Correlations between the two-/seven-point SFs (top (red dots) and bottom (blue dots) panels, respectively) and physical quantities related to stellar feedback. Left: the sum of H$\alpha$ intensities, Middle: the number of YSOs, Right: the number of {\HI} shells. Subregions including SNRs are revealed by open circles. The slope values and physical quantities are measured in the individual subregions. Spearman correlation coefficient and p-value are shown at the bottom-right corner of each panel. 
\label{fig:corr}}
\end{center}
\end{figure*}

\subsection{The Effect of Stellar Feedback on the Slope of the SF} \label{subdiscus:stellar_feedback}
Previous numerical simulations reported that stellar feedback can change the SPS slope on small scales \citep{walker2014, grisdale2017}. Based on those theoretical expectations, in the SMC, we investigated whether there are any correlations between the slope values of the two-/seven-point SFs and the physical quantities related to stellar feedback, such as the sum of H$\alpha$ intensity values, the number of YSOs, and the number of {\HI} shells in the SMC, in order to probe if stellar feedback can alter the shape of the SF (i.e., its slope). Stellar wind is also one of the important sources of stellar feedback, and it can affect {\HI} turbulence. For the SMC, however, observational data for assessing the effect of stellar wind are insufficient; thus, it is not considered in this work.

In the top panels of Figure~\ref{fig:corr}, there is no correlation between the two-point SF slope values and the three physical quantities (the sum of H$\alpha$ intensity values, the number of YSOs, and the number of {\HI} shells). These results are overall in agreement with what previous studies found in the two-point SF analysis. \cite{nestingen-palm2017} investigated whether SF slope values vary between central and outer star-forming regions in the SMC, and they found no difference. \cite{szotkowski2019} also compared the two-point SF slope map of the SMC with the star formation surface density map, which is generated by combining the MCELS H$\alpha$ data \citep{smith1999} and the {\it Spitzer} 24 $\mu$m data \citep{gordon2011}, but they did not find a clear correlation. All these results suggest that the two-point SFs for the SMC are not changed by small-scale turbulence driven by stellar feedback.

Contrary to the relatively uniform slope values of the two-point SFs across the SMC, the slope values of the seven-point SFs significantly change in the SMC (Figure~\ref{fig:sub7pt}). The bottom panels of Figure~\ref{fig:corr} show clear correlations between the slope values of the seven-point SFs and the three physical quantities. Since we used the extinction-uncorrected H$\alpha$ data, it is imperative to consider the potential impact of dust extinction on our results. Active star-forming regions with strong H$\alpha$ emission are likely to be more affected by dust extinction than the outer regions of the SMC with diffuse H$\alpha$ emission. It is therefore expected that the sum of the extinction-corrected H$\alpha$ intensity values in active star-forming regions could be more shifted toward the higher values along the x-axis of the plots in the left column of Figure~\ref{fig:corr}. Consequently, we still expect to see the correlation between the sum of H$\alpha$ intensity values and the seven-point SF slopes, indicating that the main results remain unchanged.

Interestingly, the correlation coefficient value (0.93) between the slope values of seven-point SFs and the number of {\HI} shells is relatively high, compared to other correlation coefficient values (0.72 and 0.83) related to the sum of H$\alpha$ intensities and the number of YSOs. Considering that 1) the potential energy sources for the expansion of {\HI} shells are suggested to be stellar winds and/or SN and 2) $\sim$50 pc size of the $l_{\rm SF}$ is somewhat larger than the other spatial scales (a few pc size) governed by YSOs \citep{reipurth1997}, particularly, both SNs and {\Hii} regions, which affect a few tens pc scales \citep{maggi2019,lopez2014,fichtner2024}, are likely to have played a significant role in driving turbulence, producing the {\HI} shells and characteristic fluctuation scales of 34$-$84 pc. Hereafter, we assume that the characteristic scale corresponds to the turbulent driving scale. We will discuss the characteristic scale ($l_{\rm SF}$) in more detail in the next section. 

With the characteristic scales of 34$-$84 pc and the observed {\HI} velocity dispersions of 10$-$25~km~s$^{-1}$ \citep{di_teodoro2019}, the {\HI} turbulence timescales in the subregions are calculated to be 1.3$-$8.2 Myr. By comparing the turbulence timescales with the lifetimes of turbulence driving sources, we seek to determine which source influences {\HI} turbulence over longer or shorter period in the SMC, and consequently, which one predominantly governs {\HI} turbulence.

The lifetimes of YSOs and SNRs are $\lesssim$5 Myr and $\sim$1 Myr, respectively \citep{bamba2022,dib2025}, which are roughly comparable to the timescales of {\HI} turbulence. On the other hand, the lifetime of {\Hii} regions is much longer ($\sim$18.5 Myr \citealt{mckee1997,maclow2002}) than the {\HI} turbulence timescales, suggesting that {\Hii} regions can continuously sustain {\HI} turbulence over the characteristic scales (34$-$84 pc).

However, these timescale comparisons have some caveats. First, the {\HI} turbulence timescales are lower limit values because the {\HI} velocity dispersions that we used include multiple components: (i) intrinsic velocity dispersion caused by both thermal and turbulent motions, (ii) geometrical broadening, (iii) instrumental broadening, (iv) systematic large scale shear and/or rotation, and (v) blending of velocity components \citep{di_teodoro2019,stewart2022,bhattacharjee2024}. Therefore, the true intrinsic velocity dispersion is lower, implying longer turbulence timescales. Second, the lifetimes of YSOs, SNRs, and {\Hii} regions are usually derived for conditions in the MW and may vary in the low-metallicity environment of the SMC. Given these uncertainties, it is not easy to identify the main driving source based solely on the timescale comparisons.

\begin{figure*}[!htbp]
\begin{center}
\includegraphics[width=1.00\textwidth]{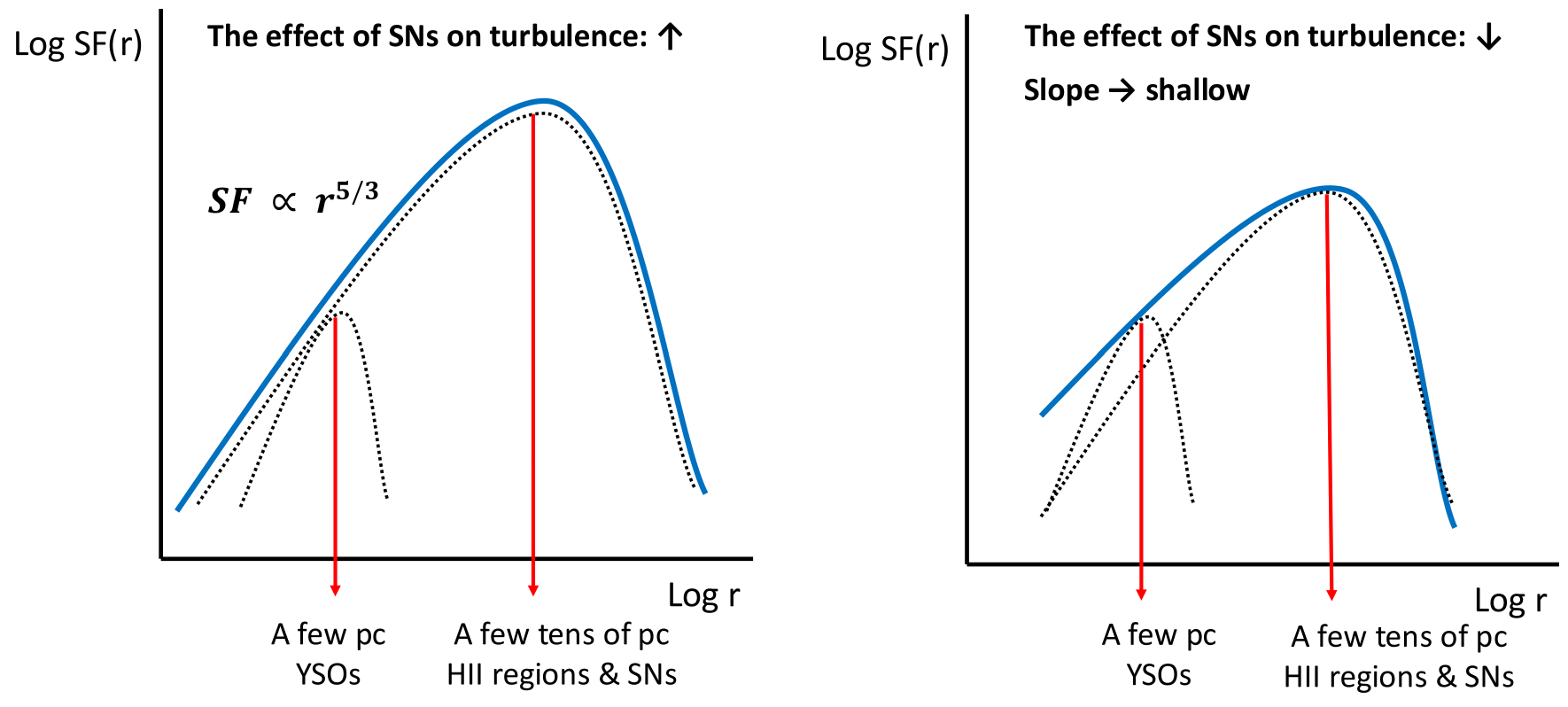} %
\caption{Cartoon illustrating the relative difference in the degree of the impact of different driving sources on the {\HI} turbulence. Black dotted lines indicate SFs originated from individual energy sources. The combined SF is shown in a blue solid line. {\Hii} regions and SNs can affect {\HI} turbulence at a spatial scale of a few ten pc, while YSOs can affect {\HI} turbulence at a spatial scale of a few pc. Left: in the case that the effect of SNs on {\HI} turbulence is dominant, the slope value of the seven-point SF is close to $\sim$1.6, as expected by \cite{cho2010}. Right: in the case that the effect SNs on {\HI} turbulence is not dominant, compared to the contribution of YSOs, the slope value of the seven-point SF becomes small. 
\label{fig:slope_cartoon}}
\end{center}
\end{figure*}

Assuming that compared to the amount of mechanical energy injected by {\Hii} regions and YSOs, SNs can inject relatively huge energy into the {\HI} gas at a spatial scale of a few tens pc, the slope value of the seven-point SF could be primarily governed by the effect of SN explosions. In particular, under the subsonic Mach number circumstances, {\HI} turbulence derived by SN explosions makes the slope value close to the slope value ($\sim$1.6) of the Kolmogorov spectrum \citep{cho2010}. For example, Subregions 12 and 19 with supernova remnants (SNRs) \citep{maggi2019} and many {\HI} shells show slope values ($\sim$1.8 and $\sim$1.7, respectively), which are close to $\sim$1.6. On the other hand, if the relative contribution of SNs to the turbulence is small, the slope value could be smaller due to an increase of the power on small scales caused by the effect of YSOs (the right panel of Figure~\ref{fig:slope_cartoon}). Although there is a difference in the magnitude of the impact of three different driving sources ({\Hii} regions, YSOs, SNs) on {\HI} turbulence, it is obvious that stellar feedback can alter the slopes of SFs. 

\subsection{{\HI} turbulence properties: The Scale Length of The Small-scale Fluctuations} \label{subsec:discus_scale_length}
In the seven-point SF analysis for the subregions of the SMC, the break feature is found between $\sim$34 pc and $\sim$84 pc across all subregions (Section~\ref{subsec:subsmc}). This indicates that the $l_{\rm SF}$ varies in the individual subregions of the SMC. Based on this result, outflows of YSOs, which can affect surrounding {\HI} gas on very small scales up to a few pc \citep{arce2010ApJ}, are unlikely to generate the $l_{\rm SF}$ of the SMC. Instead, SNs and {\Hii} regions are more likely to drive {\HI} turbulence on scales of a few tens of pc \citep[e.g.,][]{szotkowski2019}, inducing the formation of the small-scale structures at the size of $l_{\rm SF}$ and likely operating everywhere in the SMC. In particular, the $l_{\rm SF}$ of the small-scale structures is similar to the sizes of {\HI} shells \citep{hatzidimitriou2005} and {\HI} filaments \citep{ma2023}.

In the SMC, there are 509 {\HI} shells in total \citep{hatzidimitriou2005}, and these shells are widely spread over the SMC, as shown in Figure~\ref{fig:ha_yso_shell}. The mean and median values of the radius of {\HI} shells are 94 pc and 72 pc, respectively. These values are slightly larger than the $l_{\rm SF}$ of the SMC, but the size of {\HI} shells could be one of the potential candidates related to the $l_{\rm SF}$. As discussed in Section~\ref{subdiscus:stellar_feedback}, SNs seem to contribute mainly to the formation of {\HI} shells, and thus SNs may affect the determination of $l_{\rm SF}$ in the SMC.

A recent study with the GASKAP-{\HI} data found complex {\HI} filamentary structures spanning the entire galaxy \citep[][]{ma2023}. Furthermore, recent theoretical works have reported that turbulence plays a role in shaping ISM filamentary structures such as the width or the elongation of filament \citep{federrath2016, xu2019}. Based on these observations and theoretical results, we probe whether there is a connection between the $l_{\rm SF}$ and spatial scales of the {\HI} filamentary structures. Thus, we also performed the seven-point SFs with the GASKAP-$W_{\rm HI}$ map containing only filamentary structures (see details of results in Appendix~\ref{app_filament}). Interestingly, the $l_{\rm SF}$ ($\sim$42~pc to $\sim$109~pc) estimated using the GASKAP-{\HI} filament map is in good agreement with the $l_{\rm SF}$ ($\sim$34~pc to $\sim$84~pc) measured using the GASKAP-$W_{\rm HI}$ map. Since the length of the {\HI} filamentary structures is at least longer than 120 pc \citep{ma2023}, regarding the range of the $l_{\rm SF}$, the widths of {\HI} filaments or separation distances between the {\HI} filaments are more likely to be connected to the $l_{\rm SF}$ of the SMC. 

\cite{ma2023} found that the {\HI} filaments tend to align with the magnetic field, suggesting that, together with {\HI} turbulence, the magnetic field also plays a key role in shaping the {\HI} filaments of the SMC. However, further discussion about the relationship between turbulence, {\HI} filaments, and the magnetic field of the SMC is beyond the scope of this paper.

\subsection{{\HI} Turbulence Properties: Sonic Mach Number} \label{subsec:mach_num}

As shown in Figure~\ref{fig:mach_num}, sonic Mach numbers in most of the subregions of the SMC are less than one, which are overall in good agreement with Mach numbers estimated in previous works \citep{burkhart2010,gerrard2023}. This result indicates that the {\HI} gas in the SMC is subsonic or transonic, which is consistent with the fact that the {\HI} gas in the SMC is dominated by the warm neutral medium (WNM) ($n$, $T_{k}$) = (0.1$-$1~cm$^{-3}$, 4000$-$8000~K) \citep{mckee1977, wolfire2003, jameson2019, mcclure-griffiths2023}. Previous studies reported that the SMC has a low fraction ($\sim$10$-$20\%) of cold neutral medium (CNM) ($n$, $T_{k}$) = (7$-$70~cm$^{-3}$, 25$-$250~K) \citep{dickey2000, jameson2019, dempsey2022, mcclure-griffiths2023}.

It is interesting that some subregions with relatively high $M_{s}$ ($>$ 0.6) are found at the northwest part of the SMC. With our current analysis in this work, however, it is somewhat difficult to conclude which mechanism induces higher sonic Mach numbers in the northwest region of the SMC. Further investigation (e.g., using simulations) is needed to identify the dominant physical processes (e.g., interactions with the LMC and MW) driving the high Mach numbers in the northwest region of the SMC.

\section{Summary and conclusions} \label{sec:sum}

We investigated the {\HI} turbulent properties of the SMC to determine what physical mechanisms drive turbulence and at what scales. Using the high-resolution $W_{\rm HI}$ map from the GASKAP survey, we performed a statistical analysis of {\HI} turbulence with two- and seven-point SFs. In particular, the seven-point SF enables us to probe small-scale turbulence by removing large-scale fluctuations. 

Analysis of the entire {\HI} disk of the SMC reveals that while the two-point SF is relatively linear, the seven-point SF shows the break feature a scale of $\sim$50 pc. This suggests that the {\HI} turbulence in the SMC is composed of large- and small-scale variations. In particular, the identification of the break feature in the SMC was enabled by the high spatial resolution of the GASKAP data and the seven-point SF analysis.


To investigate variations in turbulent properties across the SMC, we analyzed 34 subregions of the SMC. The two-point SFs tend to be linear with slopes of $\sim$1.1 - 1.6, in good agreement with previous work \citep{szotkowski2019}, except for two regions (subregions 28 and 34). In contrast, the seven-point SFs show break features at spatial scales of 34 - 84 pc, and their slopes vary widely from $\sim$0.3 to $\sim$1.8.

The different behaviors between the two- and seven-point SFs suggest that distinct physical mechanisms govern the {\HI} turbulent properties at small- and large-spatial scales. Relatively uniform slopes with the linear trend in the two-point SFs indicate that there is dominant large-scale structures across the SMC, likely induced by gravitational interactions with the LMC and the MW. On the other hand, small-scale turbulence is thought to be driven by local processes. We found strong correlations between the seven-point SF slopes and physical quantities (H$\alpha$ intensity, the number of YSOs, the number of {\HI} shells, and SNRs) related to stellar feedback. However, the two-point SFs have no such correlations. This suggests that stellar feedback mainly affects small-scale turbulence in the SMC.

The average characteristic scale ($l_{\rm SF}$) of $\sim$50 pc is physically significant. This scale may be connected to the shape of {\HI} structures, such as the width or separation distance of filaments, or the radius of {\HI} shells. 

Sonic Mach numbers calculated using the results of seven-point SFs show that the {\HI} gas in all subregions are subsonic or transonic, which is consistent with the fact that the {\HI} gas of the SMC is dominated by the WNM. 

In future studies, the statistical analysis method with multi-point SFs can be applied to other systems (e.g., the LMC, the Magellanic Stream and Bridge, the MW) observed in the GASKAP survey, and even to other nearby extragalactic sources, in order to study details of {\HI} turbulent properties.

\acknowledgments
We gratefully acknowledge J. R. Dawson for valuable comments and discussion. B.L. acknowledges support by the National Research Foundation of Korea (NRF), grant Nos. RS-2022-NR069020. This research was partially funded by the Australian Government through an Australian Research Council Australian Laureate Fellowship (project number FL210100039) to NMc-G. C.~F.~acknowledges funding provided by the Australian Research Council (Discovery Projects DP230102280 and DP250101526), and the Australia-Germany Joint Research Cooperation Scheme (UA-DAAD). A.~S.~acknowledges support from the Australian Research Council's Discovery Early Career Researcher Award (DECRA, project~DE250100003). 
This scientific work uses data obtained from Inyarrimanha Ilgari Bundara/the Murchison Radio-astronomy Observatory. We acknowledge the Wajarri Yamaji People as the Traditional Owners and native title holders of the Observatory site. CSIRO's ASKAP radio telescope is part of the Australia Telescope National Facility (\url{https://ror.org/05qajvd42}). Operation of ASKAP is funded by the  Australian Government with support from the  National Collaborative Research Infrastructure Strategy. ASKAP uses the resources of the Pawsey Supercomputing Research Centre. Establishment of ASKAP, Inyarrimanha Ilgari Bundara, the CSIRO Murchison Radio-astronomy Observatory and the Pawsey Supercomputing Research Centre are initiatives of the Australian Government, with support from the Government of Western Australia, Australia and the Science and Industry Endowment Fund, Australia. This paper includes archived data obtained through the CSIRO ASKAP Science Data Archive, CASDA (\url{http://data.csiro.au})

%

\facilities{ASKAP}


\software{
Astropy \citep{astropy:2013, astropy:2018},  
Matplotlib \citep{Hunter:2007}, 
NumPy \citep{harris2020array}, 
SciPy \citep{2020SciPy-NMeth},
}




\clearpage

\appendix
\section{Three-, Four-, Five, Six-point Structure Functions} \label{app_multi_sf}

\begin{figure*}[!htbp]
\begin{center}
\includegraphics[width=1.00\textwidth]{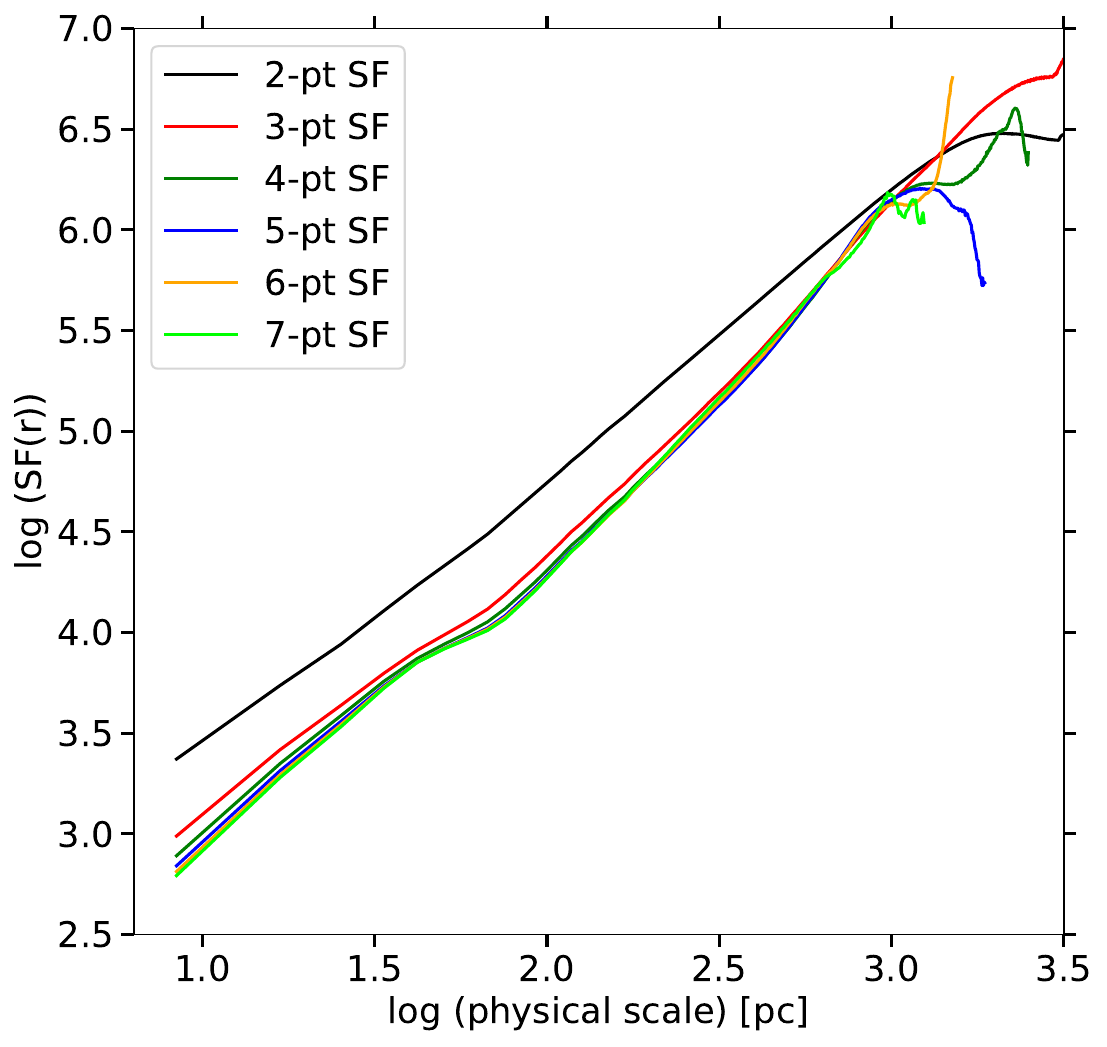} 
\caption{Results of the multi-point SFs for the entire $W_{\rm HI}$ map of the SMC.  
\label{fig:st234567_whole}}
\end{center}
\end{figure*}

\begin{figure*}[!htbp]
\begin{center}
\includegraphics[width=1.00\textwidth]{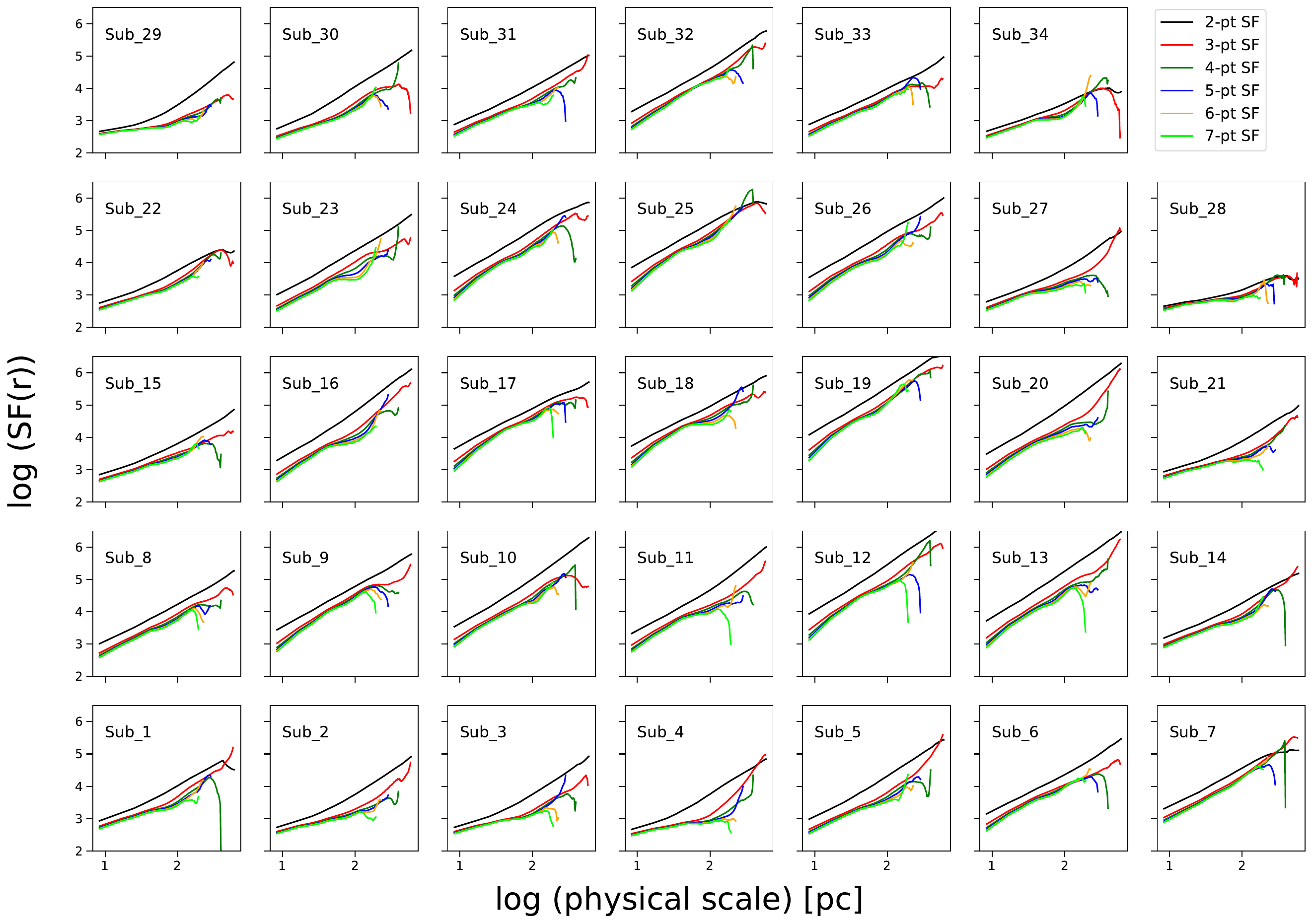} 
\caption{Results of the multi-point SFs for the 34 subregions of the SMC. 
\label{fig:st234567_sub}}
\end{center}
\end{figure*}

In addition to the two- and seven-point SFs, we also calculated three-, four-, five, and six-point SFs. These are defined by the following equations:

\begin{equation}
SF_{3pt}(r)~=~\frac{1}{3}<\lvert I(\bold{x} - \bold{r}) - 2I(\bold{x}) + I(\bold{x} + \bold{r})\rvert^2>,
\label{eq:st3pt}
\end{equation}

\begin{equation}
SF_{4pt}(r)~=~\frac{1}{10}<\lvert I(\bold{x} - \bold{r}) - 3I(\bold{x}) + 3I(\bold{x} + \bold{r}) - I(\bold{x} + 2\bold{r})\rvert^2>,
\label{eq:st4pt}
\end{equation}

\begin{equation}
    \begin{aligned}
        SF_{5pt}(r)~=~\frac{1}{35}<\lvert I(\bold{x} - 2\bold{r}) - 4I(\bold{x} - \bold{r}) + 6I(\bold{x})
        -4I(\bold{x} + \bold{r}) + I(\bold{x} + 2\bold{r})\rvert^2>,
    \end{aligned}
\label{eq:st5pt}
\end{equation}

\begin{equation}
    \begin{aligned}
        SF_{6pt}(r)~=~\frac{1}{126}<\lvert I(\bold{x} - 2\bold{r}) - 5I(\bold{x} - \bold{r}) + 10I(\bold{x}) - 10I(\bold{x} + \bold{r}) + 5I(\bold{x} + 2\bold{r}) - I(\bold{x} + 3\bold{r})\rvert^2>,
    \end{aligned}
\label{eq:st6pt}
\end{equation}
respectively\citep{cho2019,SetaFederrath2024}. Figure~\ref{fig:st234567_whole} shows the results of these SFs, together with the results of two- and seven-point SFs, for the entire $W_{\rm HI}$ map of the SMC. In addition, the multi-point SFs for the 34 subregions of the SMC are shown in Figure~\ref{fig:st234567_sub}. As recent studies using the multi-point SFs have reported \citep{SetaEtAl2023,SetaFederrath2024}, our work shows a sign of the convergence in the SF slope as the number of points in the SF increases. At large-scales, however, we can see large variations (e.g., sharp drops or rapid increases), caused by the limited map size and the small number of data points used in the SF calculations.

\section{Effect of the noise of the integrated {\HI} intensity map on the SF calculations} \label{app_noise}
\begin{figure}[!htbp]
\begin{center}
\includegraphics[width=1.00\textwidth]{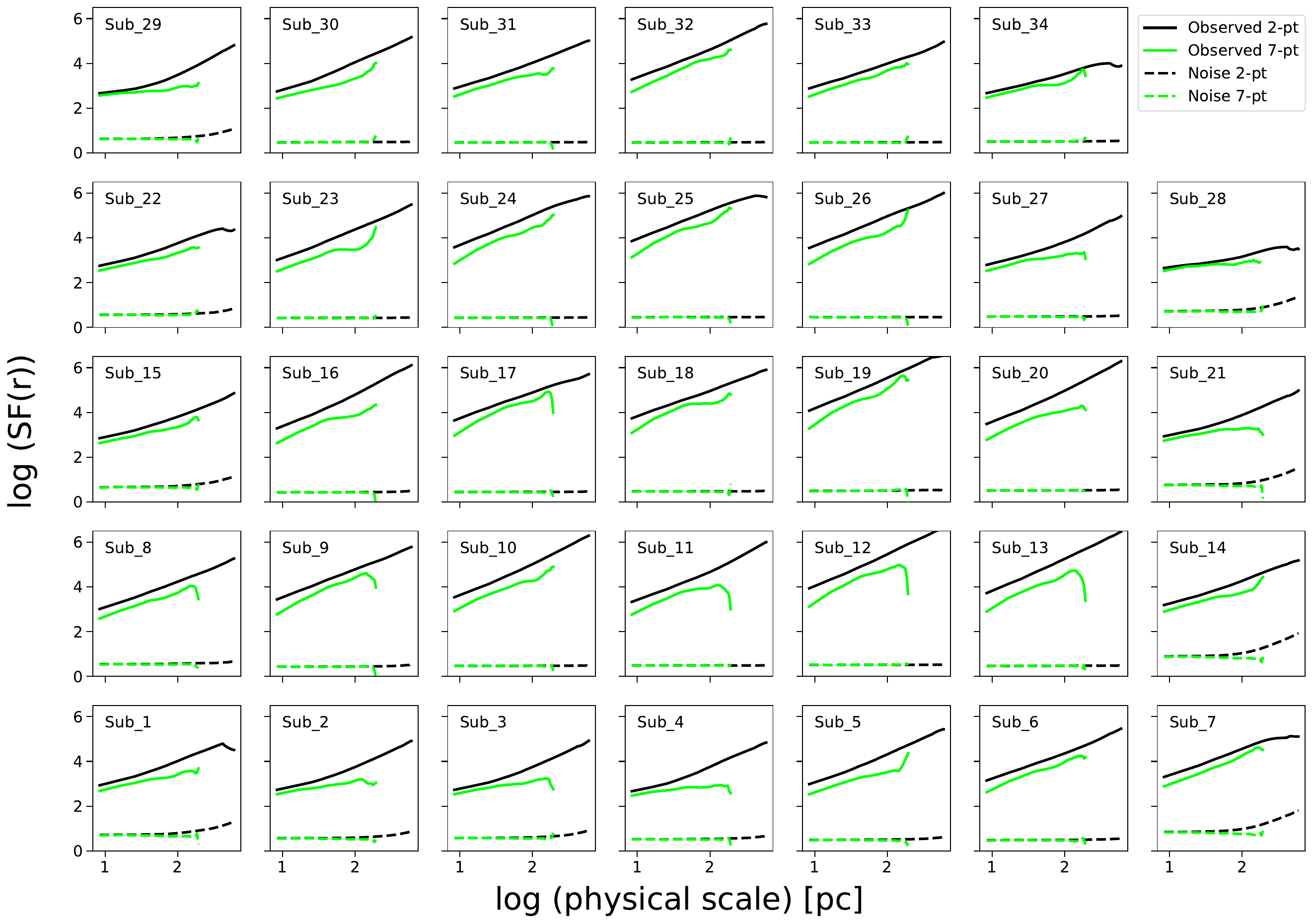} %
\caption{The observed and noise SFs (solid and dashed lines) for 34 subregions. The two- and seven-point SFs are shown in black and Green colors, respectively.
\label{fig:sf_error}}
\end{center}
\end{figure}

In Figure~~\ref{fig:wholesf}, ~\ref{fig:sub2pt}, and ~\ref{fig:sub7pt}, the observed SFs consist of the {\HI} emission SFs and the noise SFs \citep[e.g.,][]{green1993}. To probe the noise contribution to the observed SFs, we performed the SF calculations for an rms noise distribution of the integrated {\HI} map. Figure~\ref{fig:sf_error} shows noise SFs (dashed lines) together with observed SFs (solid lines). Although the noise SFs should ideally be flat, some subregions (e.g., Subregions 1 and 14) near the edge of the observing field show that two-point SFs of the noise are a non-flat shape because the edge of the observing field is very noisy, compared to the inner area of the observing field. The amplitude of noise SFs is at least two to four orders of magnitude lower than that of the observed SFs (Figure~\ref{fig:sf_error}). Therefore, the contribution of noise SFs is negligible in the observed SFs.

\begin{figure}[!htbp]
\begin{center}
\includegraphics[width=1.00\textwidth]{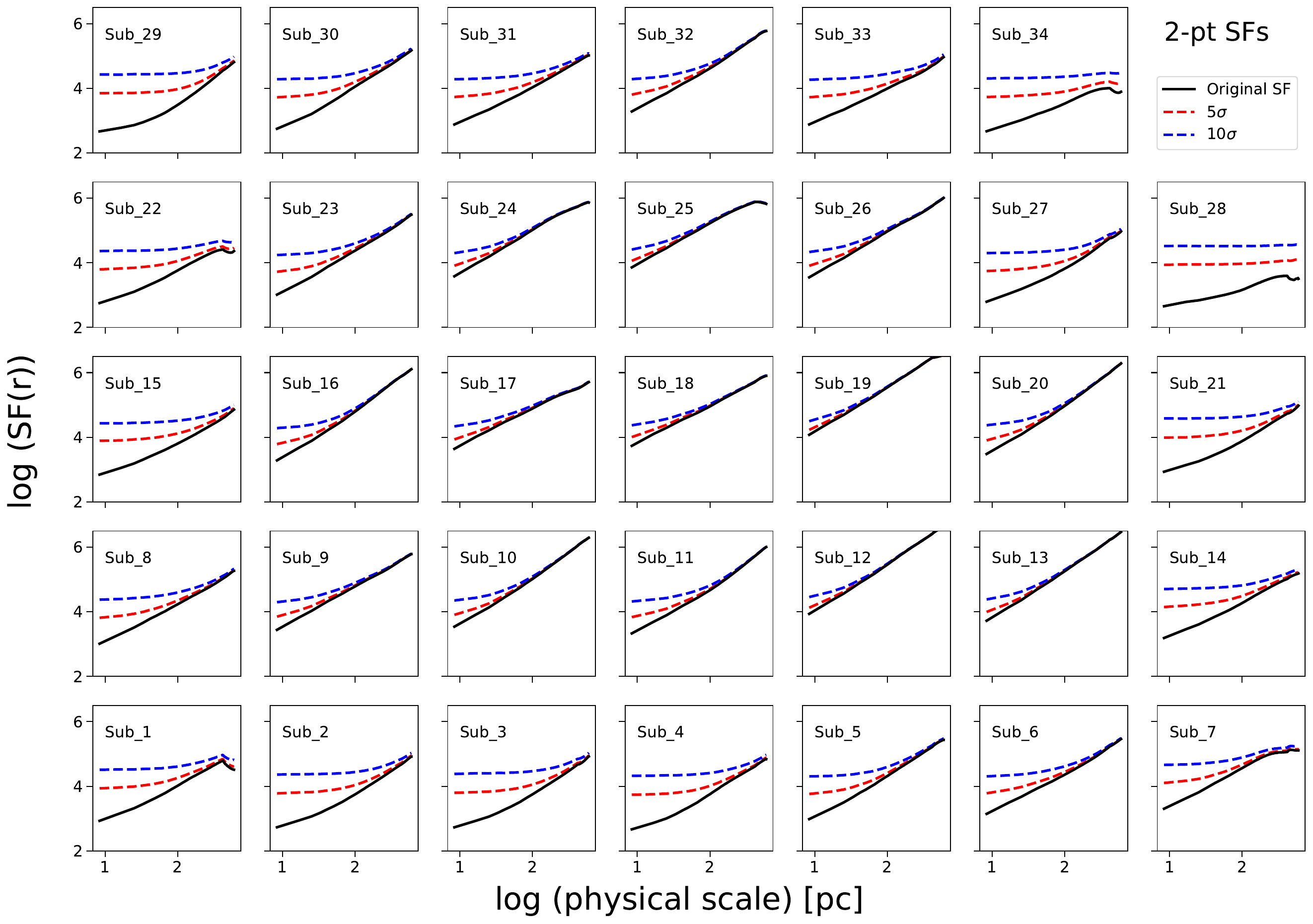} %
\caption{Two-point SFs with different noise levels. Original SFs are shown in black solid lines. Meanwhile, SFs with 5$\sigma$ and 10$\sigma$ are indicated by red and blue dashed lines, respectively.  
\label{fig:sf_2pt_noise}}
\end{center}
\end{figure}

\begin{figure}[!htbp]
\begin{center}
\includegraphics[width=1.00\textwidth]{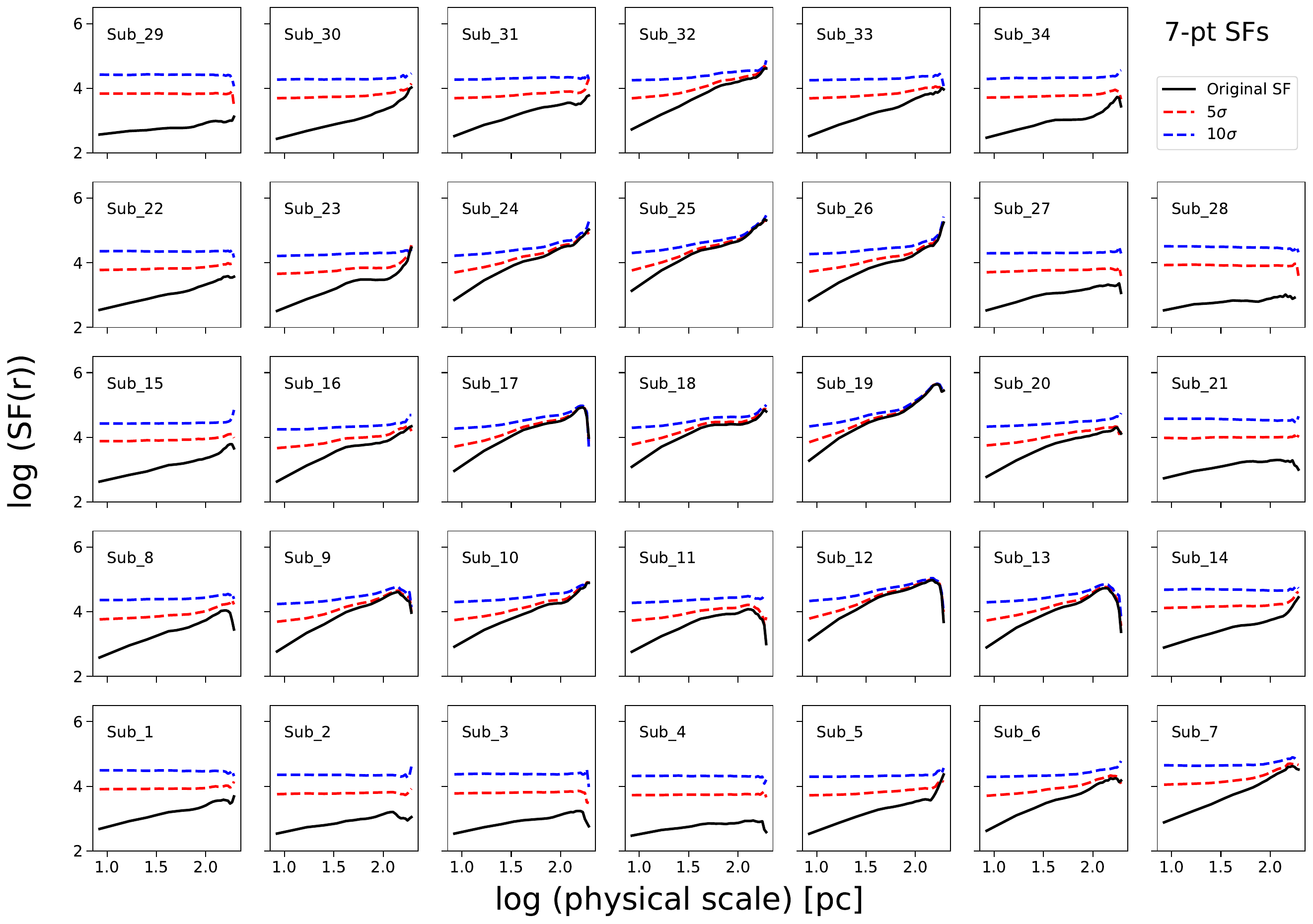} %
\caption{Seven-point SFs with different noise levels. Original SFs are shown in black solid lines. Meanwhile, SFs with 5$\sigma$ and 10$\sigma$ are indicated by red and blue dashed lines, respectively.
\label{fig:sf_7pt_noise}}
\end{center}
\end{figure}

We also investigated how different noise levels affect the slope of SF. For this examination, we create new $W_{\rm HI}$ maps with higher noise levels (5$\sigma$ and 10$\sigma$) by performing the Monte Carlo simulation. As we conducted in the error estimation of SFs (Section~\ref{sec:met}), for each noise level, 1000 $W_{\rm HI}$ maps are constructed. In these Monte Carlo simulations, we adopted standard deviation values determined by 5$\sigma$ and 10$\sigma \times \sqrt{N_{chan}} \times \Delta v$, respectively. Then, for each noise level, we randomly selected one intensity map from 1000 simulated maps and performed two- and seven-point SF calculations. 

As seen in Figure~\ref{fig:sf_2pt_noise} and ~\ref{fig:sf_7pt_noise}, the shape of SFs tends to become flat as the noise level increases. Particularly, the slope of SFs on small scales becomes shallow. This experiment indicates the fact that it is possible for subregions with low S/N ratios to have a relatively shallow slope of SFs. However, the contribution of the 1$\sigma$ noise level to the SF calculations can be negligible, as shown in Figure~\ref{fig:sf_error}. Based on those examinations of the effect of the S/N on the slope of SFs, we conclude that the slope variation of the seven-point SFs is not due to the effect of the S/N.

\section{{\HI} filament intensity map and its seven-point Structure Functions} \label{app_filament}
\begin{figure*}[!htbp]
\begin{center}
\includegraphics[width=1.0\textwidth]{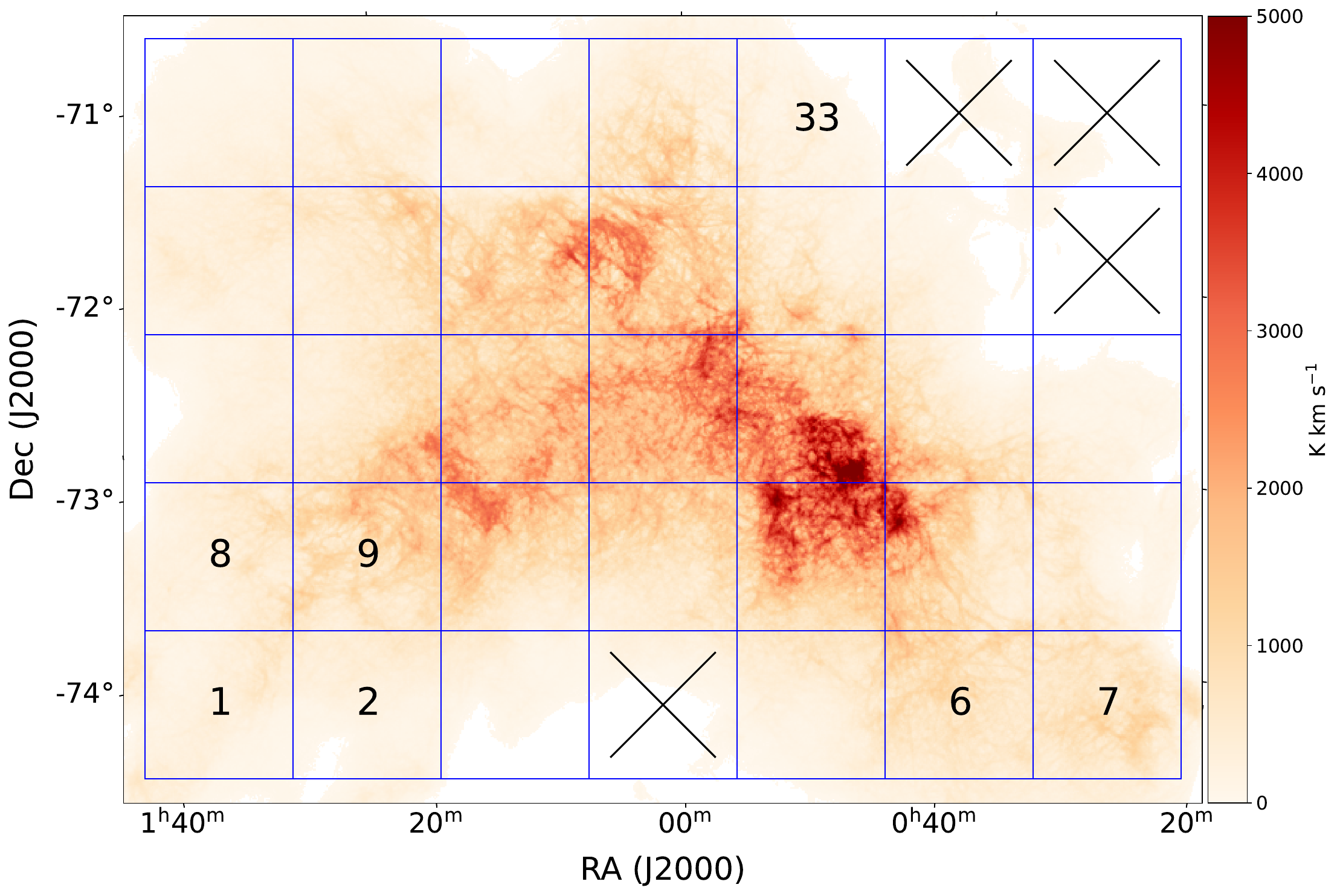}
\caption{{\HI} filament integrated intensity map of the SMC. Pixels (white) with S/N~$<$~5 and at the outside of the filamentary structures are masked. Subregions 4, 28, 34, and 35 (marked with a large cross) are excluded from the seven-point SF calculations, as more than 50\% of the total pixels in each subregion have a low S/N ratio ($<$5) or do not belong to the {\HI} filamentary structures. \label{fig_app:himap_filament}}
\end{center}
\end{figure*}

\begin{figure*}[!htbp]
\begin{center}
\includegraphics[width=1.00\textwidth]{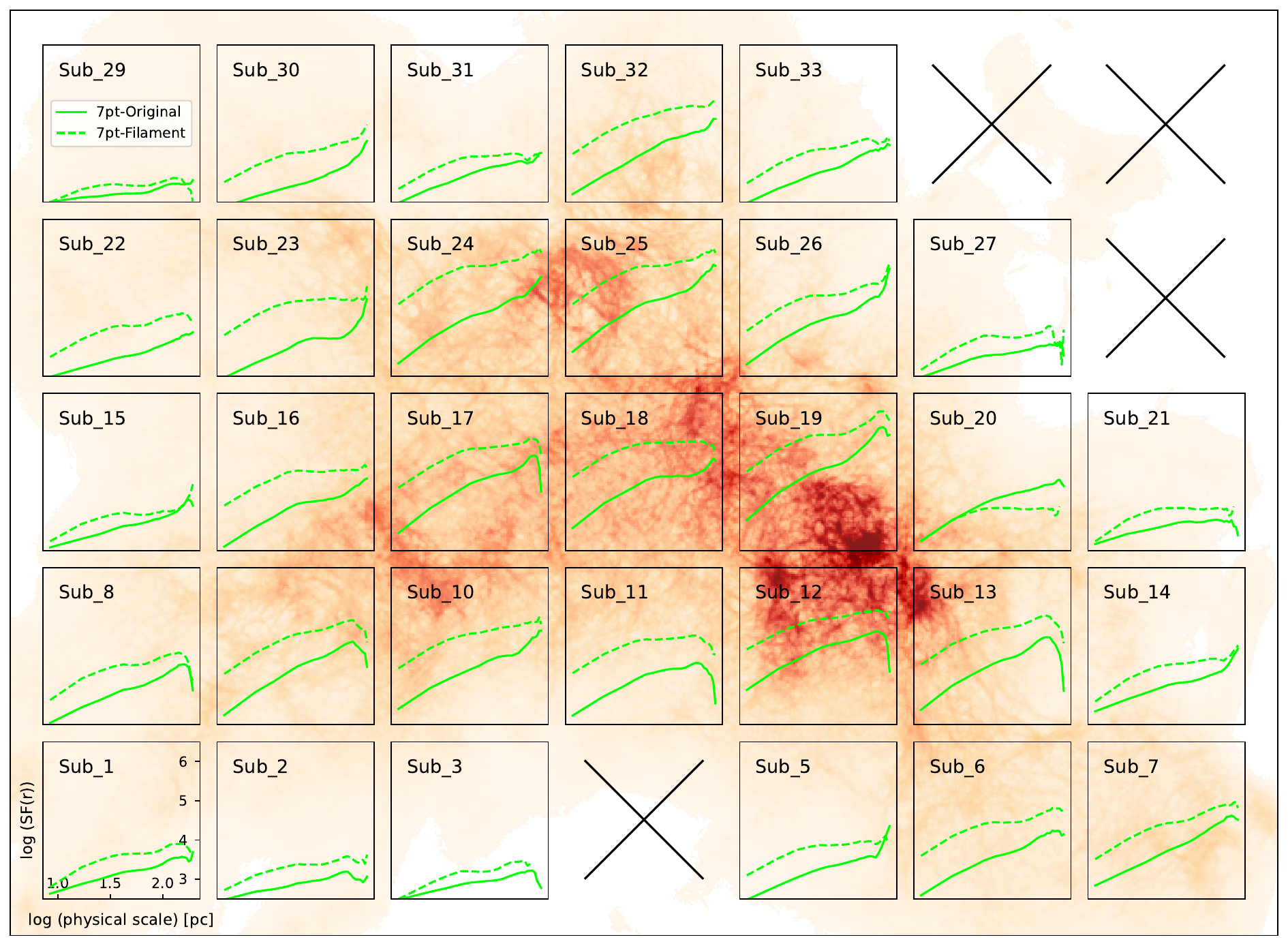} %
\caption{Seven-point SFs of the original {\HI} intensity map (solid line) and the {\HI} filament intensity map (dashed line). 
\label{fig_app:sf_7pt_filament}}
\end{center}
\end{figure*}

Using the 3D data cube containing location information of {\HI} filaments \citep{ma2023} as a mask, we could select specific pixels belonging to the {\HI} filamentary structures in each channel of the GASKAP-{\HI} cube, and then we generated the {\HI} filament intensity map using the masked GASKAP-{\HI} cube (Figure~\ref{fig_app:himap_filament}). We refer the reader to Section 2.2 of \cite{ma2023} for detailed descriptions of how to identify the {\HI} filaments in the SMC. Figure~\ref{fig_app:sf_7pt_filament} shows results of seven-point SFs in the individual regions of the {\HI} filament intensity map.

\clearpage
\bibliographystyle{aasjournal}
\bibliography{arxiv_hi_tur_smc}




\end{document}